\documentclass[a4paper, 11pt,reqno]{amsart} 
\usepackage{graphicx}
\usepackage{tikz} \usetikzlibrary{arrows,shapes} \usepackage{amssymb,
  amsthm, amsmath, amstext} \usepackage[utf8]{inputenc}
\usepackage[T1]{fontenc} \usepackage{lineno} \usepackage{cases}
\usepackage{natbib} 
\usepackage{mathrsfs}
\usepackage{dsfont}

\usepackage[dvips, ps,all]{xy}%

\usepackage{xcolor}

\newcommand{\ud}{\,\mathrm{d}} \newcommand{\mb}{\mathbf}
\newcommand{\point}{\,\cdot\,}
\newcommand{\dimens}{d}
\newcommand{\simpl}{\mb{ S}_{\dimens}}
\newcommand{\Ang}{\mathbf{w}} \newcommand{\Angrand}{\mathbf{W}}
\newcommand{\ang}{w} 
 
\newcommand{\wei}{\mathit{p}} 
\newcommand{\cdf}{\emph{cdf}}
\newcommand{\Mu}{\boldsymbol{\mu}}

\newcommand{\expmeas}{\lambda}

\newcommand{\pexc}{\zeta}
\newcommand{\Pexc}{\boldsymbol{\zeta}}
\newcommand{\prob}{\mathbf{P}}
\newcommand{\thres}{v}
\newcommand{\bthres}{\mb{\thres}}
\newcommand{\fthres}{u}
\newcommand{\bfthres}{\boldsymbol{u}}
\newcommand{\margpar}{\chi}
\newcommand{\dmpar}{\psi}

\newcommand{\datatype}{\kappa}

\newcommand{\DM}{\textsc{DM}}

\newcommand{\MCMC}{\textsc{MCMC}}

\DeclareMathOperator{\diri}{diri}
 
\DeclareMathOperator{\PRM}{PRM}
\theoremstyle{plain}

\theoremstyle{remark} 
\newtheorem*{notation*}{Notation}

\def\keywords{\vspace{.5em} {\textit{Keywords}:\, \relax%
  }}

\makeatletter
\newcommand{\authorfootnotes}{\renewcommand\thefootnote{\@fnsymbol\c@footnote}}%
 \makeatother

\begin{document}

\title{Combining regional
  estimation and historical floods: a multivariate semi-parametric
  peaks-over-threshold model with censored data}

\date{\today} 





\maketitle

\begin{center}
   

\authorfootnotes
  Anne Sabourin \textsuperscript{1}, 
Benjamin Renard \textsuperscript{2}
\end{center}
\par\bigskip
\textsuperscript{1} Institut Mines-Télécom, Télécom ParisTech, CNRS LTCI \\
37-38, rue Dareau, 75014 Paris, FRANCE\\
\texttt{anne.sabourin@telecom-paristech.fr}

\par\bigskip
\textsuperscript{2}
  Institut National de Recherche en Sciences et Technologies pour
  l'Environnement et l'Agriculture, 
  Centre de Lyon \\ 
  5 rue de la Doua - CS70077, 
  69626 VILLEURBANNE Cedex, France


\par\bigskip
\thispagestyle{empty}
\pagestyle{empty}


\begin{abstract}
  The estimation of extreme flood quantiles is challenging due to
  the relative scarcity of extreme data compared to typical target return periods.
  Several approaches have been developed over the years to face this challenge, 
  including regional estimation and the use of historical flood data. This paper investigates
  the combination of both approaches using a multivariate peaks-over-threshold model, 
  that allows estimating altogether the intersite dependence structure and
  the marginal distributions at each site. The joint distribution of extremes at several sites 
  is constructed using a semi-parametric Dirichlet Mixture model. The existence of partially missing and
  censored observations (historical data) is accounted for within a data
  augmentation scheme. This model is applied to a case study involving four catchments in Southern France, 
  for which historical data are available since 1604. The comparison of marginal estimates from four versions
  of the model (with or without regionalizing the shape parameter; using or ignoring historical floods)
  highlights significant differences in terms of return level estimates. Moreover, the availability of historical data
  on several nearby catchments allows investigating the asymptotic
  dependence properties of extreme floods. Catchments
  display a  a significant amount of asymptotic dependence, calling for adapted multivariate statistical models. 
\end{abstract}

\keywords{Multivariate extremes; censored data; semi-parametric
  Bayesian inference; mixture models; reversible-jump algorithm}

\section{Introduction}
\label{intro}

Statistical analysis of extremes of  uni-variate hydrological time
series is a relatively well chartered problem. Two main representations can be used in the context of extreme
value theory \citep[\emph{e.g.}][]{Madsen1997b, coles2001introduction}: block maxima (typically, annual maxima) can be modeled 
using a Generalized Extreme Value (GEV) distribution \citep[see \emph{e.g.}][]{Hosking1985}, while flood peaks
over a high threshold (POT) are commonly modeled with 
a Generalized Pareto (GP) distribution \citep[see \emph{e.g.}][]{Hosking1987,davison1990models,Lang1999}. 
One major issue in at-site flood frequency analysis is related to 
data scarcity \citep{neppel2010flood}: as an illustration, most of the recorded flood time
series in France are less than $50$ years long, whereas flood return periods of
interest are typically well above $100$ years. Moreover, an additional challenge arises if one is interested in 
multivariate extremes at several locations. A complete understanding of the joint behavior of
extremes at different locations requires to model their dependence structure as well. While there exists a 
multivariate extreme value theory \citep[\emph{e.g.}][]{coles1991modeling,DeHaan1998}, its practical application is
much more challenging than with standard univariate approaches.

\subsection{Regional estimation} \label{intro:regional}
In order to address the issue of data scarcity in at-site flood frequency analysis, 
hydrologists have developed methods to jointly use data from several sites: 
this is known as Regional Frequency Analysis (RFA) \citep[\emph{e.g.}][]{Hosking1997,Madsen1997,Madsen1997c}.
The basis of RFA is to assume that some parameters governing the
distributions of extremes remain constant at the
regional scale  \citep[see \emph{e.g.} the 'Index Flood' approach of ][]{Dalrymple1960}.
All extreme values recorded at neighboring stations can hence be used to estimate the
regional parameters, which increases the number of available data.

The joint use of data from several sites induces a technical difficulty: the spatial dependence
between sites has to be modeled. A common assumption has been to simply ignore 
spatial dependence by assuming that the observations recorded simultaneously at different sites are
independent, which is often unrealistic 
\citep[see][for appraisals of this assumption]{Stedinger1983,Hosking1988,Madsen1997}.
An alternative approach uses elliptical copulas to describe spatial
dependence \citep{Renard2007,Renard2011}.
While this approach allows moving beyond the spatial independence assumption, it is not fully satisfying. 
Indeed, such copula models are not compatible with multivariate extreme value theory \citep{resnick1987extreme,Resnick07,Beirlant04}.
This  may alter uncertainty assessments about regional parameters (in particular for shape parameters) and, in turn,  
about extreme quantiles. In this context, using a dependence model compatible with multivariate extreme value theory is of interest. 

\subsection{Historical data} \label{intro:historical}
Beside regional analysis methods, an alternative way to reduce uncertainty is
to take into account historical flood records to complement the
systematic streamflow measurements over the recent period
\citep[see \emph{e.g.}][]{Stedinger1986,OConnel2002,Parent2003,Reis2005,naulet2005flood,neppel2010flood,Payrastre2011}.
This results in a certain amount of
censored and missing data, so that any likelihood-based inference ought
to be conducted using a censored version of the
likelihood function. Also, in a regional POT context, some
observations may not be concomitantly extreme at each location,
so that the marginal GP distribution does not apply to them.
A `censored likelihood' inferential framework for extremes
has been introduced to take into account such observations
\citep{smith1994multivariate, ledford1996statistics, smith1997markov}.
The information carried by partially censored data is likely to be all the
more relevant in a multivariate, dependent context, where information
at one  well gauged location can be transferred  to poorly measured ones. 

\subsection{Multivariate modeling}
The family of admissible dependence structures between extreme events
is, by nature, too large to be fully described by any parametric model 
(see further discussion in section \ref{subsec:dependence model}). 
For applied purposes, it is common to restrict the dependence model to a
parametric sub-class, such as, for example, the  Logistic model and its
asymmetric and nested extensions
\citep{gumbel1960distributions,coles1991modeling, stephenson2003simulating,stephenson2009high}.
The main practical advantage is that the censored
versions of the likelihood are readily available, but parameters are
subject to non-linear constraints  and structural modeling choices
have to be made \emph{a priori}, \emph{e.g.}, by 
allowing only bi-variate or tri-variate dependence between closest neighbors. 
An alternative to parametric modeling  is to resort to `semi-parametric' 
mixture models (some would say `non-parametric' because it can
approach any dependence structure): the distribution function characterizing the
dependence structure is written as a weighted average of an
arbitrarily large number of 
simple parametric components. This allows keeping the practical advantages of a parametric
representation while providing a more flexible model. 

\subsection{Objectives: Combining historical data and regional analysis} \label{intro:combining}
Our aim is to combine regional analysis and
historical data
by modeling altogether the marginal distributions and the dependence
structure of excesses above large thresholds at neighboring
locations with partially censored data. 
Combined historical/regional approaches have been explored by a few authors \citep{Tasker1987,Tasker1989,Jin1989,Gaume2010}.
This paper builds on this previous work and extends it to a multivariate POT context,
where each $d$-variate observation corresponds to concomitant streamflows recorded at $d$ sites.
This is to be compared with the multivariate annual maxima approach, where each $d$-variate observation
corresponds to componentwise annual maxima that may have been recorded during distinct extreme episodes.

In this paper, a multivariate POT model is implemented in order to combine 
regional estimation and historical data. This model is used to investigate two scientific questions. 
Firstly, the relative impact of regional and historical information on marginal quantile estimates at each site is investigated.
Secondly, the existence of historical data describing exceptional flood events at several nearby catchments provides an unique 
opportunity to investigate the nature and the strength of intersite dependence at very high levels
(which would not be possible using short series of systematic data only).


Multivariate POT modeling is implemented in a Bayesian, semi-parametric context.
The dependence structure is described using a Dirichlet Mixture ( \DM\ ) model.
The \DM\  model was first introduced by
\cite{Boldi_Davison07}, and  its reparametrized version
\citep{sabourinNaveau2012}  allows for  Bayesian inference  with a
varying   number of mixture components.  A complete description of the model and of the reversible-jump
Markov Chain Monte-Carlo ( \MCMC\ ) algorithm used for inference with non censored data is given in
\cite{sabourinNaveau2012}. 
The adaptation of the inferential framework to the case of partially censored and missing
data is fully described from a statistical point of view in a
forthcoming paper \citep{sabourin14censoring}\footnote{
  preprint available online at 
  \texttt{http://perso.telecom-paristech.fr/\~{}sabourin/}}
One practical advantage of this  mixture model is that no additional structural modeling
choice needs to be made, which allows to cover an arbitrary wide range
of dependence structures. 
In this work, we aim at modeling the multivariate distribution of
$\dimens = 4$ locations. However, the methods presented here are theoretically
valid in any dimension, and computationally realistic in moderate
dimensions (say $\dimens\le 10$ ).

The remainder of this paper is organized as follows: the dataset under
consideration is described in Section~\ref{sec:dataDescript}, and a
multivariate declustering scheme is proposed to handle temporal
dependence. Section ~\ref{sec:model} summarizes the main features of
the multivariate POT model and describes the inferential algorithm. In Section
\ref{sec:results}, the model is fitted to the data and results are
described. Section~\ref{sec:discussion} discusses the main limitations 
of this study and proposes avenues for improvement,
while Section~\ref{sec:conclusion} summarizes the main findings of this study.

\section{Hydrological data }\label{sec:dataDescript}
\subsection{Overview}
\begin{figure}[hbtp]
  \begin{center}
    \includegraphics[scale=0.7]{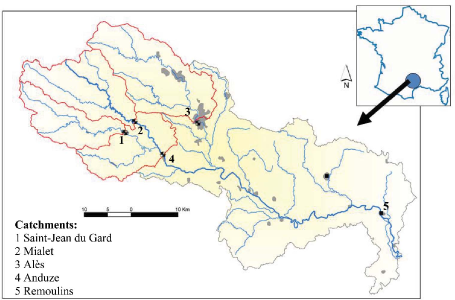}
    \caption{Hydrological map of the area of the Gardons, France
      \citep{neppel2010flood}}
    \label{fig:map}
  \end{center}
\end{figure}

The dataset under consideration  consists of
discharge 
recorded in the area of the `Gardons', in the south of France.  Four
catchments (Anduze, 540 $km^2$, Alès, 320 $km^2$, Mialet, 219 $km^2$, and Saint-Jean,154 $km^2$) are considered. They are
located relatively close to each other (see Figure~\ref{fig:map}).
Discharge data (in
$m^3.s^{-1}$) were reconstructed by \cite{neppel2010flood} from
systematic measurements (recent period) and historical
floods. \cite{neppel2010flood} estimated separately the marginal uni-variate
extreme value 
distributions for yearly maximum discharges, taking into account
measurement and reconstruction errors arising from the conversion of
water levels into discharge. The earliest record dates back to
$1604$, September $10^{th}$ and the latest was made in $2010$,
December $31^{st}$.  

In this work, since we are more interested in the dependence structure
between simultaneous records than between yearly maxima, we   model multivariate excesses over
threshold, and the variable of interest becomes (up to declustering)
the daily peakflow.
Of course, most of the $N = 14\,841$ daily peaklflows  are censored (\emph{e.g.}, most historical
data  are only known to be smaller than the yearly
maximum for the considered year). For the sake of simplicity, we do not take into account any
possible measurement errors. 

The geographic 
proximity of the four considered stations suggests dependence at high
levels. This is visually confirmed by the pairwise plots in Figure~\ref{fig:bivarPlots},
obtained after declustering (see Section \ref{sec:dataDescript}). 

The marginal data are classified into four different types,
numbered from $0$ to $3$: 
`$0$' denotes  missing data,   `$1$' indicate an `exact'  record. 
Data of type `$2$' are  right-censored:  the discharge  is known to be
greater than a given value. Finally, type `$3$' data are  left- and right-censored: the discharge is known
to be comprised between a lower (possibly $0$) and an upper bound. 
Most data on the historical period are of type $3$. 
In the sequel, $j \ (1\le j\le \dimens)$ denotes the location index and
$t\  ( 1\le t\le n)$ is used for the 
the temporal one.  A marginal
observation $O_{j, t}$ 
is  a $4$-uple $O_{j, t} = (\datatype_{j, t}, Y_{j, t}, L_{j, t},  R_{j, t}) \in
\{0,1,2,3\}\times \mathbb R^3$, where $ \datatype, Y, L$ and $ R$ stand
respectively for the data type, the recorded  discharge (or some
arbitrary  value if $ \datatype\neq 1$, which we denote $\texttt{NA}$), the lower bound (set to $0$ if
missing),  and the upper bound (set to $+\infty$ if missing).


\subsection{Data pre-processing: extracting cluster maxima}\label{sec:preprocess}
Temporal dependence is handled  by declustering, \emph{i.e.} by
fitting the model to cluster maxima instead
of the raw daily data. The underlying assumption is that only short term
dependence is present at extreme levels, so that excesses above high
thresholds  occur in clusters. Cluster maxima are treated as
independent data to which a model for threshold excesses may be
fitted.   For an introduction to declustering techniques, the
reader may refer to \cite{coles2001introduction} (Chap.5). 
For more details, see \emph{e.g.}
\cite{leadbetter1983extremes}, or  \cite{davison1990models} for
applications when the quantities of interest are cluster maxima. Also,
\cite{ferro2003inference} propose a method for
identifying  the optimal cluster size, after estimating  the extremal
index. However, this latter approach relies heavily on `inter-arrival
times', which are not easily available in our context of censored
data.  
In this study, we adopt a simple `run declustering' approach, following
\cite{coles1991modeling} or  \cite{nadarajah2001multivariate} :
a multivariate declustering threshold $\bthres
= (\thres_1,\dots \thres_\dimens)$ is specified (typically, $\bthres=
(300, 320, 520, 380)$ 
respectively for Saint-Jean, Mialet, Anduze and Alès),
as well as a duration $\tau$ representative of the hydrological
features of the catchment (typically $\tau = 3$ days). Following
common practice \citep{coles2001introduction}, 
the thresholds  are  chosen in regions of stability of the
maximum likelihood estimates of the marginal parameters.

In a censored data context, a marginal data $O_{j,t}$ exceeds $\thres_j$
(\emph{resp.} is below $\thres_j$) if
$\datatype_{j,t}=1$ and $Y_{j,t} >\thres_j$ (\emph{resp.} $Y_{j,t}< \thres_j$), or if
$\datatype_{j,t}\in\{2,3\}$ and $L_{j,t}>\thres_j$ (\emph{resp.}
$\datatype_{j,t}=3$ and $R_{j,t}< \thres_j)$. 
If none of these conditions holds, we say that the data point has undetermined position with
respect to the threshold. This is typically the case when some
censoring intervals intersect the declustering thresholds whereas  no
coordinate is above threshold.

A 
cluster is initiated when at least one marginal observation $O_{j, t}$ exceeds the
corresponding marginal threshold $\thres_j$.  It ends only when, during at least
$\tau$ successive days, all marginal observations are either below their
corresponding threshold, or have  undetermined position.
Let $\{ t_i \,, 1\le i\le n_\bthres\}$ be the temporal indices  of cluster starting dates. 
A  cluster maximum  $\mb C^\vee_{t_i}$ is   the
component-wise `maximum' over a cluster duration
$[{t_i},\dotsc,t_i+r]$. Its definition require special care in the
context of censoring: 
the marginal cluster maximum is  $C_{j,t_i}^\vee = \left(
  \datatype_{j,t_i}^\vee,   Y_{j,t_i}^\vee,  L_{j,t_i}^\vee,  R_{j,t_i}^\vee\right) $,  with
$Y_{j,t_i}^\vee=  \max_{t_i\le t\le t_i+r}\{  Y_{j, t}  \}$ and similar
definitions for $L_{j,t_i}^\vee,R_{j,t_i}^\vee$.
The marginal type $ \datatype_{j,t_i}^\vee$ is that of the `largest' record
over the duration. More precisely, omitting the temporal index, 
if    $Y_j^\vee >  L_j^\vee$, then   $\datatype_{j}^\vee =
1$. Otherwise, if $ L_{j}^\vee <  R_{j}^\vee $, then
$\datatype_{j}^\vee$ is set to $= 3$; otherwise, if $ L_{j}^\vee >0$, then
$\datatype_{j}^\vee = 2$ ; If none of the above holds, then the
$j^{\text{th}}$ cluster coordinate is missing and
$\datatype_j^{\vee}=0$.

\begin{figure}[hbtp]
  \centering
  \includegraphics[scale=0.38]{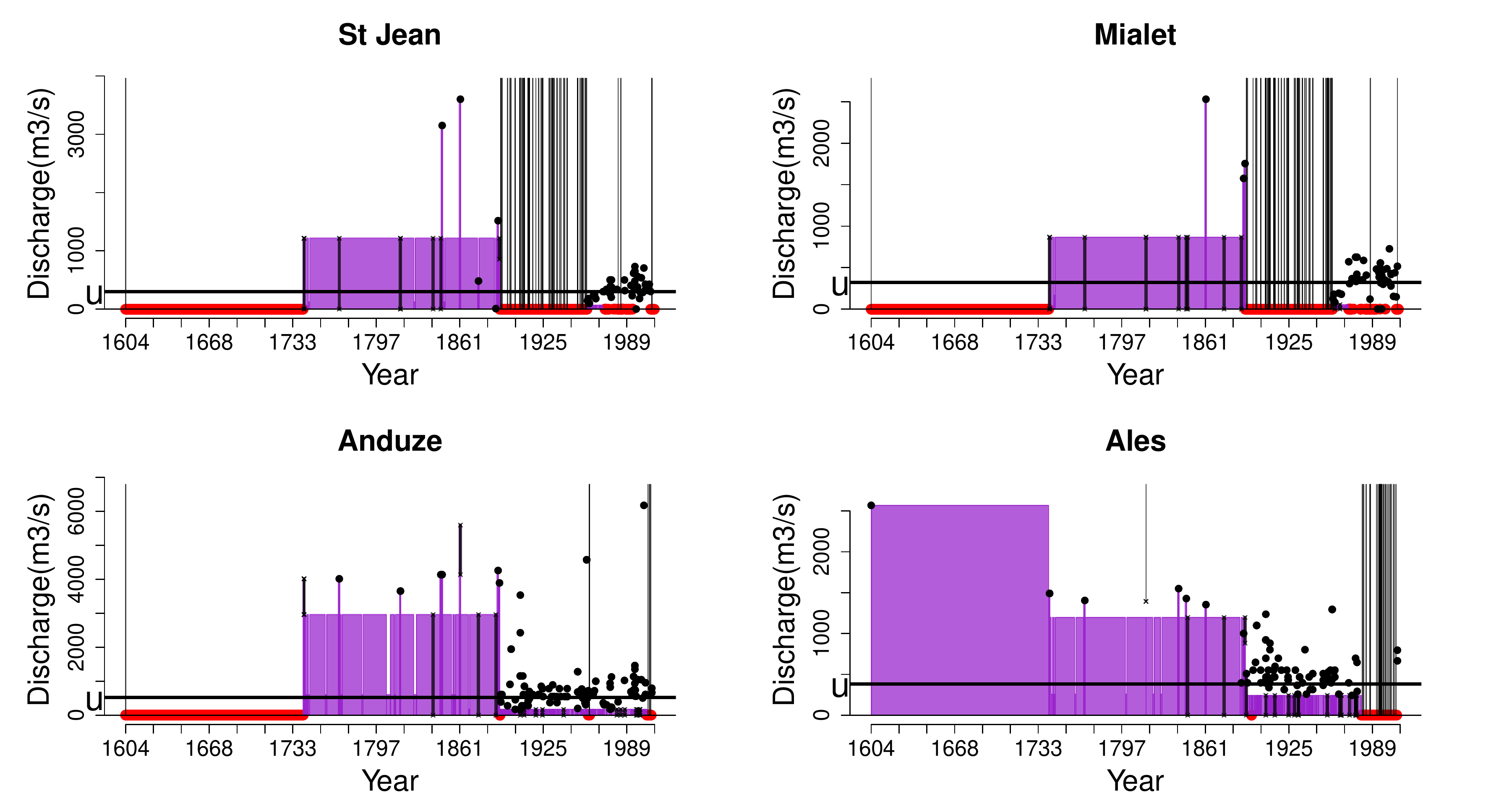}
  \caption{Extracted peaks-over-threshold at the four considered
    stations. Violet segments and areas represent data of type $2$ and
    $3$ available before declustering. Missing days are shown in red.
    Grey segments (\emph{resp.} black points) are data of type 2 and 3
    (\emph{resp. 1}) belonging to an extracted multivariate cluster. The declustering
    threshold is represented by the horizontal black line. Vertical Grey
    lines are drawn at days which are missing at the considered location
    but which belong to a cluster, due to a threshold excess at another location.}
  \label{fig:declustUni}
\end{figure}

Figure~\ref{fig:declustUni} shows the uni-variate projections of the multivariate
declustering scheme, at each location.  Points and
segments below the declustering threshold  indicate situations when
the threshold was not exceeded at the considered location but at
another one.

Anticipating Section~\ref{sec:model}, marginal cluster maxima  below
threshold are censored in the statistical analysis, so that their  marginal
types are always set to $3$, with lower bound at  zero and upper bound
at the threshold. This approach, fully described \emph{e.g.} in
\cite{ledford1996statistics}, prevents from having to estimate the
marginal distribution below threshold, which does not participate in
the dependence structure of extremes. 

After declustering and censoring below threshold, the data set is made of $n_{\bthres}= 125$ $d$-variate
cluster maxima $\{ \mb C_{t_i}^\vee, 1\le i\le n_v\}$.
The empirical mean cluster size  is $\hat\tau =1.248$, which is to
be used as a normalizing constant for  the number of inter-cluster
days. Namely, $m$ dependent 
inter-cluster  observations contribute to the likelihood as  $m/\hat\tau$
independent ones would do (see \emph{e.g.}
\cite{Beirlant04}, Chap. 10 or \cite{coles2001introduction}, Chap. 8).
As for those inter-cluster observations, $n_{\mathrm{bel}}=  7562$ data points are below thresholds and only $9$ days are
completely missing (no recording  at any location). The
remaining $n'_\bthres = 140\,674$ days are undetermined, 
and must be  taken into account in the likelihood expression. 
They can be classified into $34$
homogeneous temporal blocks  (\emph{i.e.} all the days within a given
block contain the same information), typically, between two recorded
annual maxima. The block  sizes are $n'_i (1\le
i \le 34)$, so that $\sum_{i=1}^{34} n'_i = n'_{\bthres}$. 

\begin{figure}[hbtp]
  \centering
  \includegraphics[scale=0.38]{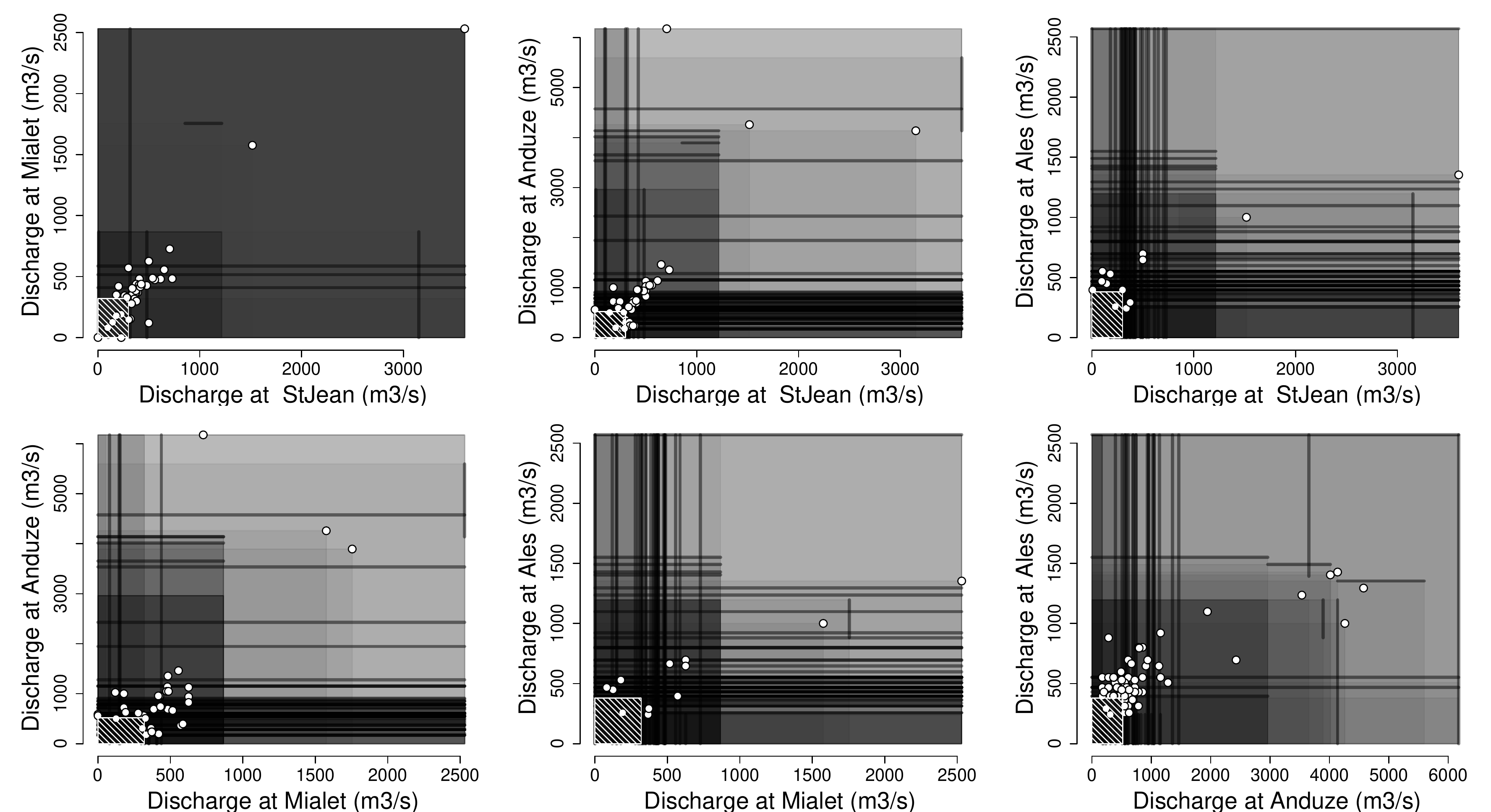}
  \caption{Bi-variate plots of the $124$ simultaneous stream-flow records
    (censored cluster maxima) at the four stations, and of the $34$ 
    undetermined data blocks defined in section~\ref{sec:preprocess},  over the whole period
    1604-2010. Points represent exact data, gray lines and
    squares respectively represent data
    for which one (\emph{resp.} two)   coordinate(s) is (are)
    censored. Data superposition is represented by increased
    darkness. The striped rectangle at the origin is the region
    where all coordinates are below  threshold. 
    \label{fig:bivarPlots}}
\end{figure}

Figure~\ref{fig:bivarPlots}  shows bi-variate plots of the extracted
cluster maxima together with undetermined blocks. Exact data are represented
by points; One coordinate missing or censored yields a segment and
censoring at both locations results in a rectangle.
The plots show  the asymmetrical nature of  the problem under
study: the  quantity of available data varies  from one pair to
another (compare, \emph{e.g.}, the number of points available
respectively for the
pair Saint-Jean/Mialet and Saint-Jean/Alès). Joint modeling
of excesses thus appears as a way of transferring  information
from one location to another.  Also, the most extreme
observations seem to occur simultaneously (by pairs): They  are more
numerous  in
the upper right corners  than near the axes,
which suggests the use of a  dependence structure model for  asymptotically
dependent data  such as
the Dirichlet mixture (see Section \ref{subsec:dependence model}).

\section{Multivariate peaks-over-threshold model}\label{sec:model}

This section provides a short description of the statistical model
used for estimating the joint distribution of excesses above high
thresholds. A more exhaustive statistical description is given in the
above mentioned forthcoming paper. 
For an overview of  statistical 
modeling of extremes in hydrology, the reader may refer \emph{e.g.} to 
\cite{katz2002statistics}. Also, \cite{davison1990models} focus on the
uni-variate case and \cite{coles1991modeling} review the most
classical multivariate extreme value models. 

\subsection{Marginal model}

After declustering,
the extracted cluster maxima are assumed to be independent from each
other. Their margins (values of the cluster maxima  at each location considered separately)
can be modeled by a    Generalized Pareto distribution above
threshold, provided that 
the latter is chosen high enough \citep{davison1990models,coles2001introduction}. 
Let $Y_{j,t_i}^\vee$ be the
(possibly unobserved) maximum water discharge at station $j$, in
cluster $i$ and let $F_j^\bthres$ the marginal cumulative distribution
function (\emph{c.d.f.}) below threshold. 
The  marginal probability of an excess above threshold is denoted  $\pexc_j
\,\;(1\le j\le d)$.
Following common practice \citep[\emph{e.g.}][]{coles1991modeling,davison1990models,ledford1996statistics},
$\pexc_j$ is identified with its empirical estimate  $\hat \pexc_j$,
which is 
obtained as   the proportion of  
intra-cluster days (after uni-variate declustering) among the
non-missing days for the considered margin and threshold.  
For $\bthres$ as above, it yields $ \Pexc \simeq (0.0021, 0.0022, 0.0022, 0.0020)$.

The marginal  models are thus 
\begin{equation*}
  \begin{aligned}
    F_j^{(\xi_j, \sigma_j)}(y) &= \prob(Y_{j,t_i}^\vee < y | \xi_j, \sigma_j)\;,  \qquad (1\le j\le d) \\
    &=\begin{cases}
      1 - \pexc_j \left(1+\xi_j\frac{y-\thres_j}{\sigma_j}
      \right)^{-1/\xi} &  (\text{if }y\ge \thres_j),   \\
      (1-\pexc_j) F_j^\bthres (y) & (\text{if }y<  \thres_j).
    \end{cases}
  \end{aligned}
\end{equation*}

The marginal parameters are gathered into a $2\dimens$-dimensional vector
$$
\margpar = \left(\log(\sigma_1),\dotsc,\log(\sigma_\dimens),
  \xi_1,\dotsc,\xi_d\right)\,,
$$
and the uni-variate \emph{c.d.f.}'s are denoted by $F_{j}^{\margpar}$.

In a context of  regional frequency analysis, it is further assumed that the shape parameter of 
the marginal GP distributions is identical for all catchments, i.e. $\xi_1=\dotsb=\xi_d$.

\subsection{Dependence structure} \label{subsec:dependence model}

In order to apply probabilistic results from multivariate extreme
value theory, it is convenient to handle Fréchet distributed variables 
$X_{j,t_i}$, so that $P(X_{j,t_i} < x) = e^{-\frac{1}{x}},\; x>0$. 
This is achieved by  defining a marginal transformation 
$$  \mathcal{T}_{j}^{ \margpar}(y) = -1/\log\left(F_j^{\margpar}(y)
\right ), $$ 
and letting $ X_{j,t_i} = \mathcal{T}_j^{\margpar}(Y_{j,t_i}) $.
The
dependence structure is then defined between  the Fréchet-transformed
data. 
One  key assumption underlying  multivariate extreme value models is
that  random vectors $\mb Y_t =
(Y_{1,t}, \dotsc,Y_{\dimens,t})$ are regularly varying
\citep[see \emph{e.g.}][]{resnick1987extreme,Resnick07,Beirlant04,coles1991modeling}.
Multivariate regular variation (MRV) can be expressed as a radial
homogeneity property of  the distribution of the largest
observations: For any region $A\subset (\mathbb{R}^+)^d$
bounded away from $0$, if we denote 
$r.A = \{\mb x\in \mathbb{R}^d : \frac{1}{r} \mb x \in A\}$, then,
for large $r_0$'s and for $r>r_0$, MRV and transformations to unit-Fréchet
imply that 
\begin{equation}
  \label{eq:MRV}
  r\, \prob(\mb X\in r.A) \underset{r_0\rightarrow \infty, r>r_0}{\sim}
  r_0 \,\prob(\mb X\in r_0.A) \,.    
\end{equation}

Switching
to a pseudo-polar coordinates system, let 
$
R = \sum_{j=1}^d X_{j}
$
denote the radius and $\Angrand = 
(\frac{X_1}{R},\dotsc,\frac{ X_d}{R})$ denote the angular
component of the Fréchet re-scaled data. In this context, $\Angrand$ is a point on
the simplex $\simpl$: $\sum_{j=1}^d W_j = 1,\, W_j\ge 0$.
Then \eqref{eq:MRV} implies that, for any angular region 
$B\subset \simpl$,
\begin{equation}
  \label{eq:limitH}
  \prob(\mb W \in B \,|\,R>r_0) \underset{r_0\rightarrow \infty}{\longrightarrow} H(B)
\end{equation}
where $H$ is the so-called `angular probability measure', \emph{i.e.}
the distribution of the angles corresponding to large radii. 
Since  in addition,
$\prob(R>r_0) \underset{r_0\rightarrow \infty}{\sim} \frac{d}{r_0}$, 
the joint behavior of large excesses is entirely determined by  $H$. 

\begin{figure}[hbtp]
  \centering
  \includegraphics[scale=0.3]{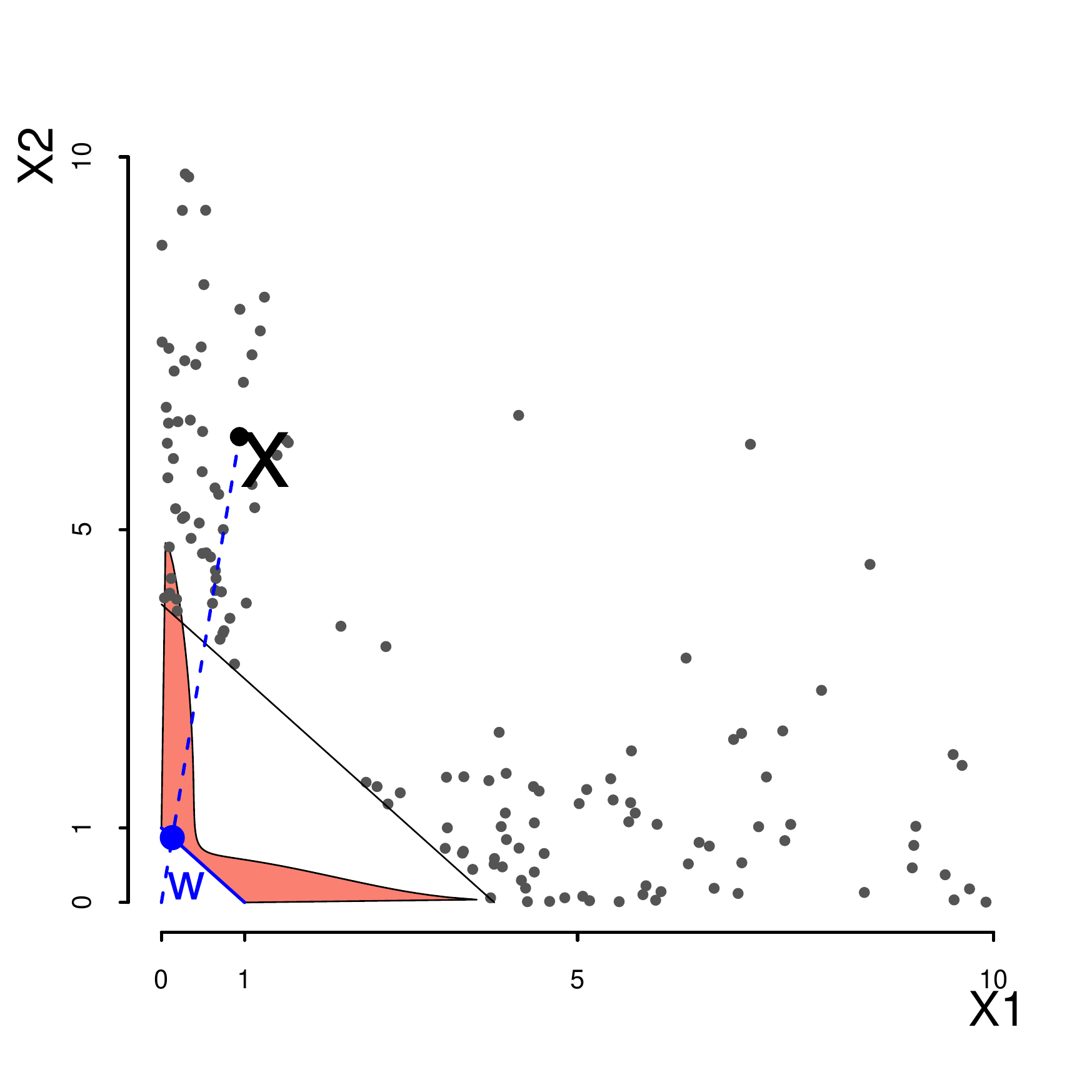}
  \includegraphics[scale=0.3]{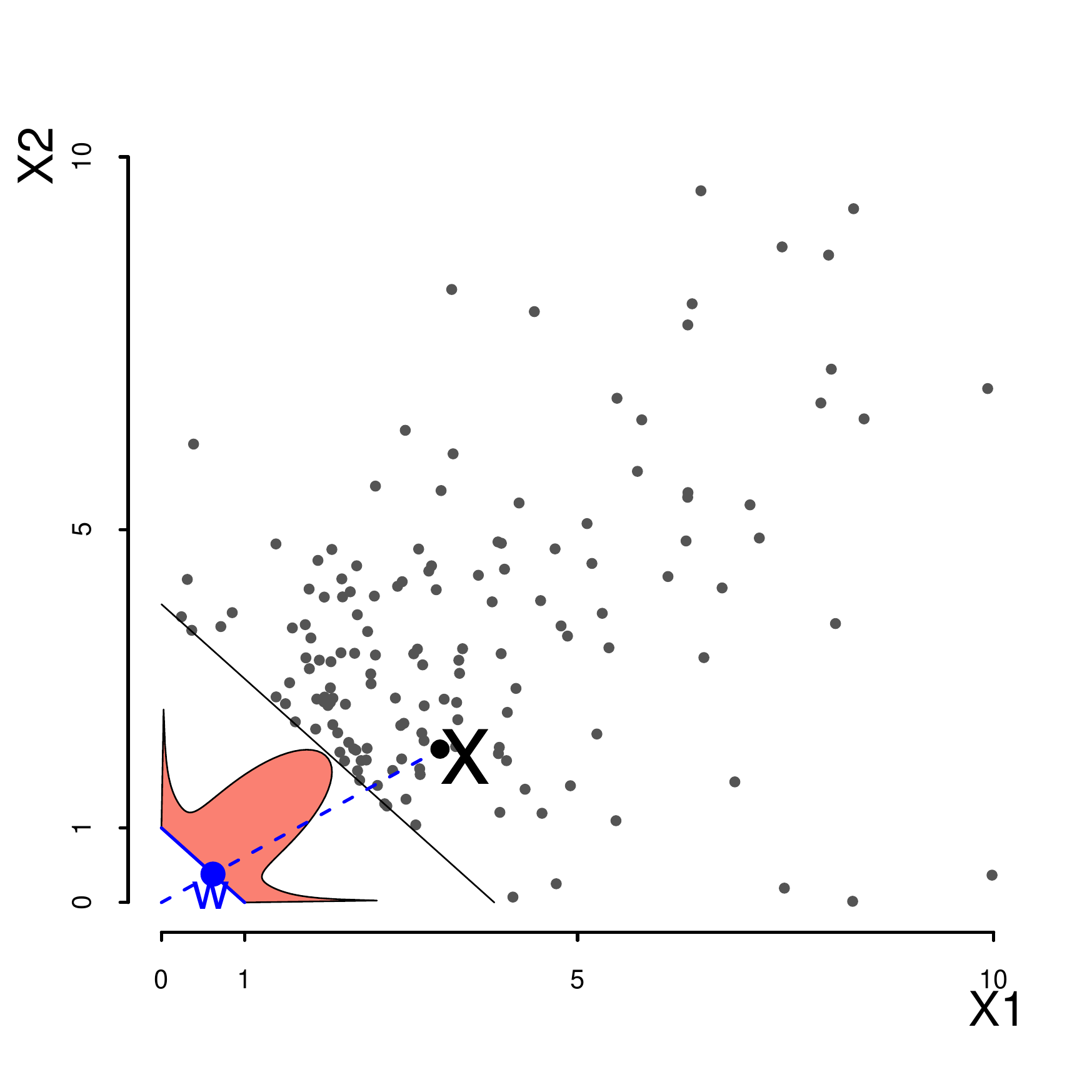}
  \caption{Two Examples of bivariate dependence structures of excesses
    above a radial threshold.\newline
    Grey points: simulated bivariate data. Pale red area: density of
    the angular distribution. Blue point: one randomly chosen angle $\mb W$, corresponding
    to the observation $\mb X$ (black point).}\label{fig:exampleH}
\end{figure}

As an illustration of this notion of angular distribution, Figure~\ref{fig:exampleH} shows two examples of simulated
bi-variate data sets, with two different angular distributions and
same   Pareto-distributed radii. $H$'s  density is represented by the pale red
area. In the
left panel, $H$ has most of its mass  near the end points of the
simplex (which is, in dimension 2,  the segment $[(1,0), (0,1)]$,
represented in blue on Figure~\ref{fig:exampleH}) and
the extremes are weakly dependent, so that events which are large  in both components
are scarce. In the limit case where $H$  is  concentrated at  the
end-points of the simplex (not shown), the pair is said to be asymptotically
independent.  In contrast, the right panel shows a case of strong
dependence: $H$ is concentrated near the middle point of the simplex
and extremes occur mostly simultaneously. 


Contrary to the limit distribution of uni-variate excesses, $H$ does
not have to belong to any particular parametric family. The only constraint on
$H$ is due to the standard form of the $X_j$'s: $H$ is a valid
angular distribution if and only if 
\begin{equation}\label{moments}
  \int_{\simpl} \ang_j \ud H(\Ang) = \frac{1}{d}\quad (1\le j\le d)\,.
\end{equation}

In this paper, $H$ is chosen in the  Dirichlet
mixture model \citep{Boldi_Davison07},  which can
approach any valid angular distribution. 
In short, a Dirichlet distribution with shape  $\nu\in \mathbb{R}^+$ and
center of mass $\Mu \in \simpl$ has density 
$$
\diri_{\nu,\Mu}(\Ang) = \frac{\Gamma(\nu)}{\prod_{j=1}^d
  \Gamma(\nu\mu_j)}
\prod_{j=1}^d \ang_j^{\nu\mu_j-1}\,.
$$
The density of a Dirichlet mixture distribution is therefore a weighted
average of Dirichlet densities. 
A parameter for a $k$-mixture is thus of the form
$$\dmpar = \left( (\wei_1,\dotsc,\wei_k), (\Mu_{\point,1},\dotsc,
  \Mu_{\point,k} ), (\nu_1,\dotsc,\nu_k)
\right),$$
with weights $\wei_m>0$, $\sum_m \wei_m =1$, which will be denoted by 
$\dmpar = \left( \wei_{1:k}, \Mu_{\point,1:k}, \nu_{1:k}
\right)$.
The corresponding
mixture density is 
$$
h_{\dmpar}(\Ang) = \sum_{m=1}^k \wei_m \diri_{\nu,\Mu_{\point,m}}(\Ang)\;.
$$
As for the moment constraint  \eqref{moments}, it  is satisfied if and only if 
\begin{equation}\label{centerMassCondition}
  \sum_{m=1}^k \wei_m \Mu_{\point,m} = \left(1/\dimens,\dotsc,1/\dimens\right)\, .
\end{equation}
In other terms, the center of mass of the
$\Mu_{\point,1:m}$'s, with weights $\wei_{1:m}$, must lie at the center of
the simplex. 

\subsection{Estimation using censored data}

Data censorship is the main  technical issue in this paper. 
This
section exposes the matter as briefly as possible. For the sake of
readability, technical details and full statistical justification 
have been  gathered in the above mentioned unpublished  paper. 


In order to account for  censored data overlapping threshold  and censored or missing components in the
likelihood expression, it is convenient to write the model in terms of
a Poisson point process, with intensity measure determined by
$H$. More precisely, after marginal standardization,
 the time series of  excesses  above large
thresholds
can be described as a Poisson point process ($\PRM$),
$$
\sum_{t=1}^{n} \mathds{1}_{\left(t, {\mb X_t}
  \right)} \sim \PRM(\ud s \times \ud \expmeas )\quad \text{on
}[0,n]\times  A_{\bfthres}\,,
$$

where $n$ is the length of the observation period,
$A_{\bfthres}$ is the `extreme' region 
on the Fréchet scale, 
$A_{\bfthres} = [0,\infty]^d \setminus [0,\fthres_1]\times \dotsb \times 
[0,\fthres_d]$, above  Fréchet thresholds  $\fthres_j
= \mathcal{T}_j^{\margpar}(\thres_j) = -1/\log(1-\Pexc_j)$. 
The notation  $\ud s$ stands for  the Lebesgue measure on $[0,1]$
and $\lambda$ is the so-called `exponent measure', which is 
related to the angular distribution's density  $h$ \emph{via} 

$$
\frac{\ud \lambda}{\ud \mb x}(\mb x) = 
d. h(\mb w) r^{-(d+1)} \qquad 
\Big( r=\sum_{j=1}^d x_j\,,\, \mb w = \mb x /r\,\Big)\,.
$$
This Poisson model has been 
widely used for
statistical modeling of extremes 
\citep{coles2001introduction,coles1991modeling,joe1992bivariate}. The
major advantage in our context is that it allows to take into account
the undetermined  data (which cannot be ascertained  to be below nor
above threshold), as they correspond to events of the kind 
$$\mb N \left\{\left[{t'_i}, {t'_i + n'_i}\right] \times  
  \left([0,\infty]^d\setminus [0, \mathcal{T}_1^\margpar(R_{1,t'_i})] \times
    \dotsc [0, \mathcal{T}_d^\margpar(R_{d,t'_i})]\right) \right\}= 0\,,
$$
where $\mb N\{\point\}$ is the number of points from  the Poisson
process in a given region. 

In our context, $h$ is a Dirichlet mixture density: $h=h_{\psi}$. Let $\theta = \left(\margpar, \dmpar\right)$ represent the parameter
for the joint model, and $\lambda_\dmpar$ be the Poisson intensity
associated with $h_{\psi}$.
The likelihood in  the Poisson model, in the absence of censoring, is 
\begin{equation}\label{eq:fullLikelihood}
  \begin{aligned}
    \mathcal{L}_{\bthres} \left( \{\mb y_t\}_{1\le t
        \le n}, \theta
    \right) 
    & \propto e^{-n\,\expmeas_{\dmpar}(A_{\bfthres})}
    \prod_{i=1}^{n_{\bthres}} \Big\{
    \frac{\ud\expmeas_{\dmpar}}{\ud \mb x} ( \mb x_{t_i})
    \prod_{ j: y_{j,t_i} > \thres_j} J_j^{\margpar}(y_{j,t_i})
    \Big\}\,. 
  \end{aligned}
\end{equation}
The $J_j^\margpar$'s are the Jacobian terms accounting for the
transformation $\mb y\to \mb x$.

The likelihood function in presence of such undetermined data and of 
censored data above threshold 
is obtained by integration of \eqref{eq:fullLikelihood} in the
direction of censorship. These integrals do not have a closed form
expression. 
In a Bayesian context, a Markov Chain Monte-Carlo
(\MCMC) algorithm is built in order to sample from the posterior
distribution, and the censored likelihood is involved at each
iteration. Rather than using numerical approximations,
whose bias may be difficult to assess,  one 
option is to use a \emph{data augmentation} framework \citep[see
\emph{e.g.}][]{tanner1987calculation, van2001art}. The main idea is to
draw the missing coordinates from their full conditional distribution
in a Gibbs-step of the \MCMC\ algorithm. Again,
technicalities  are omitted here.

\section{Results}\label{sec:results}

In this section, the multivariate extreme model with Dirichlet mixture
dependence structure is fitted to the data from the Gardons, including
all historical data and assuming a regional shape parameter. 
This regional hypothesis is confirmed (not rejected) by a likelihood
ratio test: the p-value of the $\chi^2$ statistic is $0.16$.
To assess the added value of taking into account historical data on
the one hand, and
of a regional analysis on the other hand,  inference is also made  without the
regional shape assumption and considering only the systematic measurement period
(starting from January, 1892). Thus, in total,  four model fits are performed.

For each of the four experiments, $6$ chains of $10^6$ iterations 
are run in parallel, which requires a moderate computation
time\footnote{The execution time ranged 
  from approximately  $3$h$30'$ to  $4$h$30'$ for each chain on  
  a standard processor Intel  $3.2$ GHz.}.
Using parallel chains allows to check convergence using standard stationarity and mixing tests
(\cite{heidelberger1983simulation}'s test ,
\cite{gelman1992inference}'s variance ratio  test),
available in the \textsc{R} statistical software. In the remainder of
this section, all posterior predictive
estimates are computed using the last $8\,10^5$ iterations of the
chain obtaining the best stationarity score.

\begin{figure}[hbtp]
  \centering
  \begin{center}
    Shapes   
  \end{center}

  \includegraphics[scale=0.3]{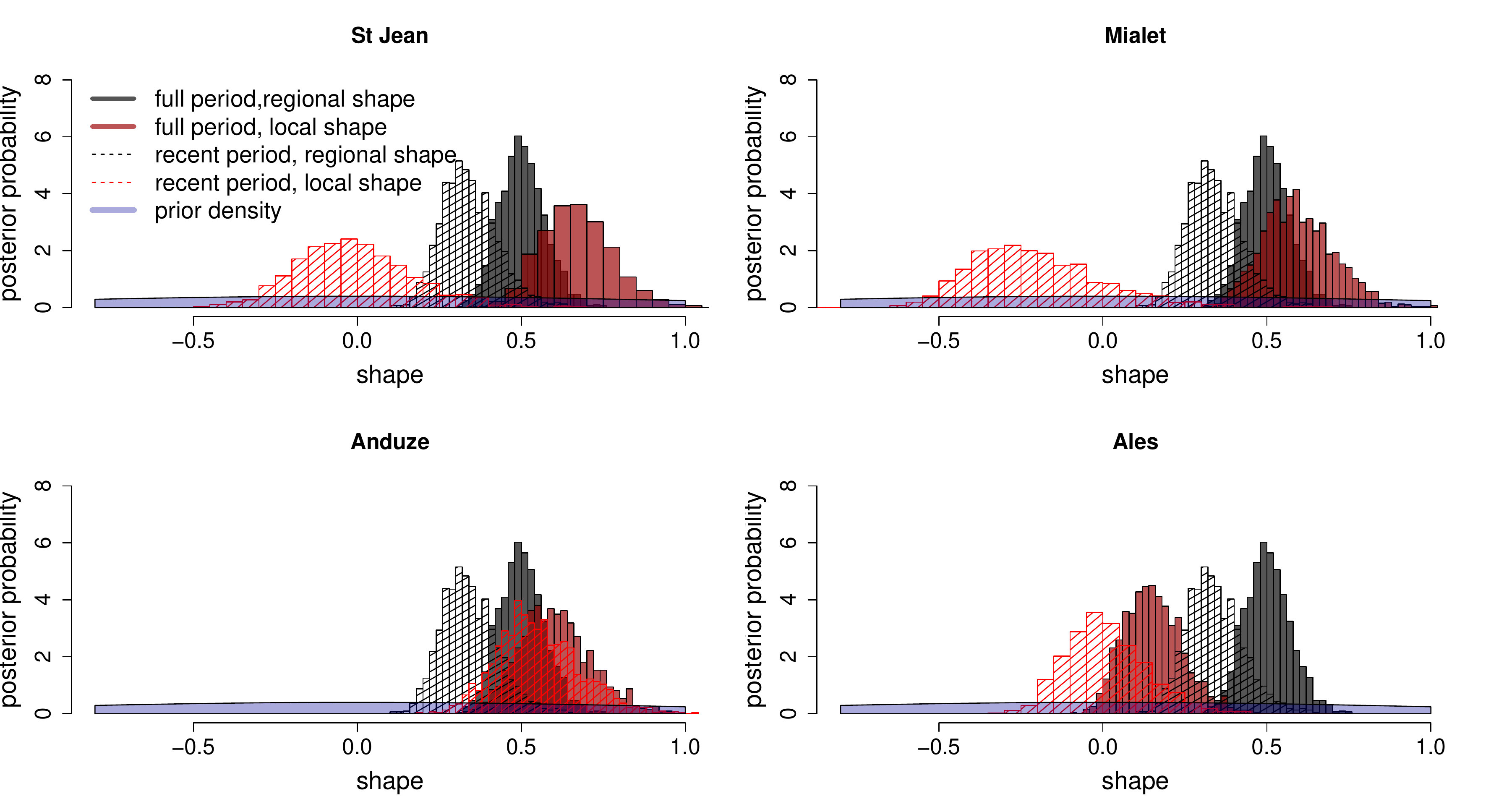}
  \line(1,0){250}
  \begin{center}
    Scales  
  \end{center}

  \includegraphics[scale=0.3]{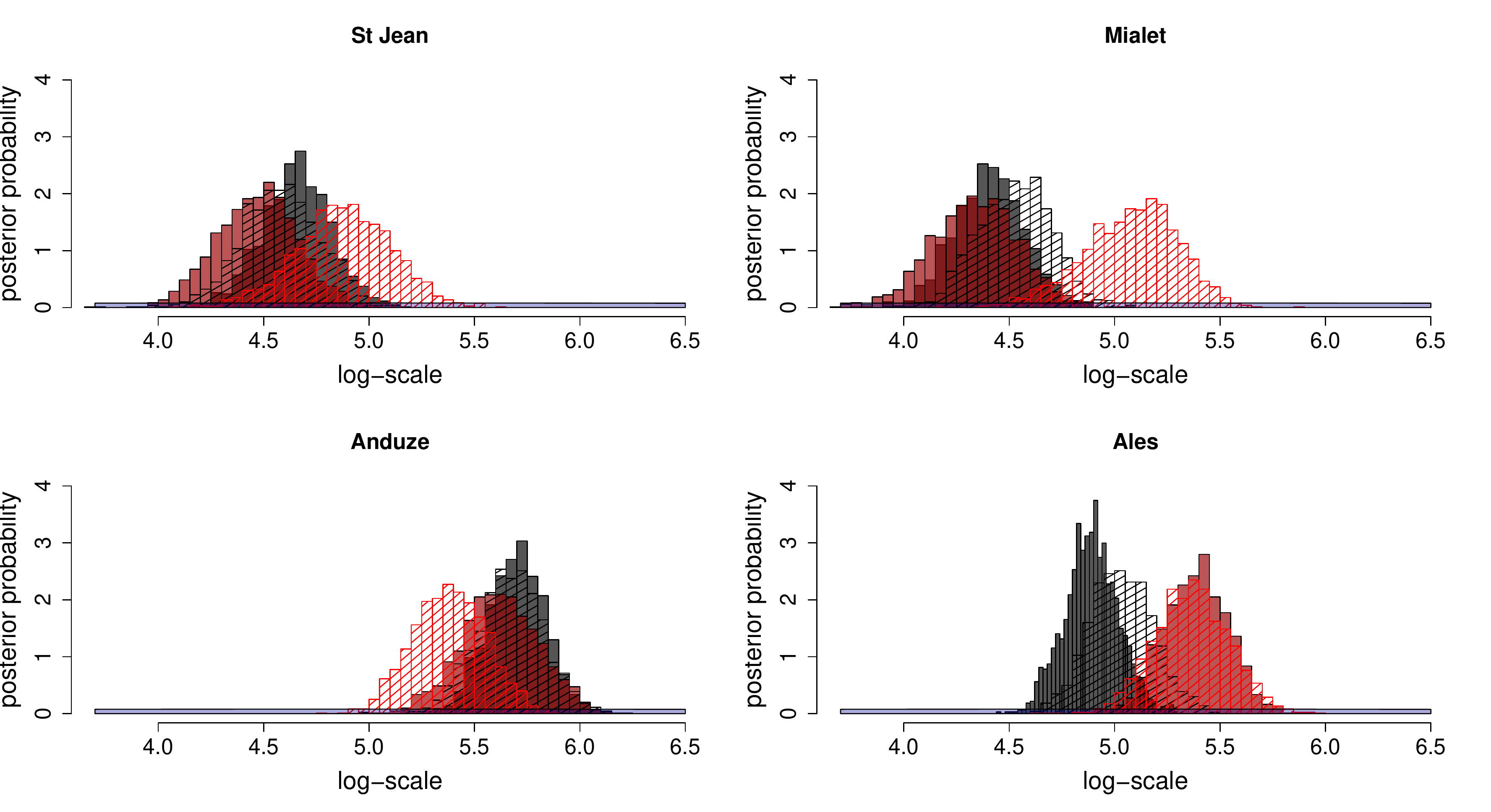}
  \caption{Prior and posterior distributions of the shape parameter (upper panel)
    and of the logarithm of the scale parameter (lower panel) at the
    four locations, estimated with or without historical data, in a
    regional framework or not. }\label{fig:posteriorHistMarg}

\end{figure}

Figure~\ref{fig:posteriorHistMarg} shows posterior histograms of the
marginal parameters, together with the prior density. The posterior
distributions are much more concentrated than the priors, indicating that
marginal parameters are identifiable in each model. Also, the shape
and scale panels are almost symmetric: a posterior distribution
granting  most weight to comparatively high  shape
parameters  concentrates on comparatively  low scales. This
corroborates
the fact that frequentist estimates of the shape and the scale
parameter are negatively correlated \citep{ribereau2011note}. 
In the regional model as well as in the local one, the posterior
variance of each parameter is  reduced when taking into account
historical data (except for the scale parameter at Anduze, for the
local model). This confirms the general fact that taking into account
more data tends to reduce the uncertainty of parameter estimates.

\begin{figure}[hbtp]
  \centering
  \includegraphics[scale=0.38]{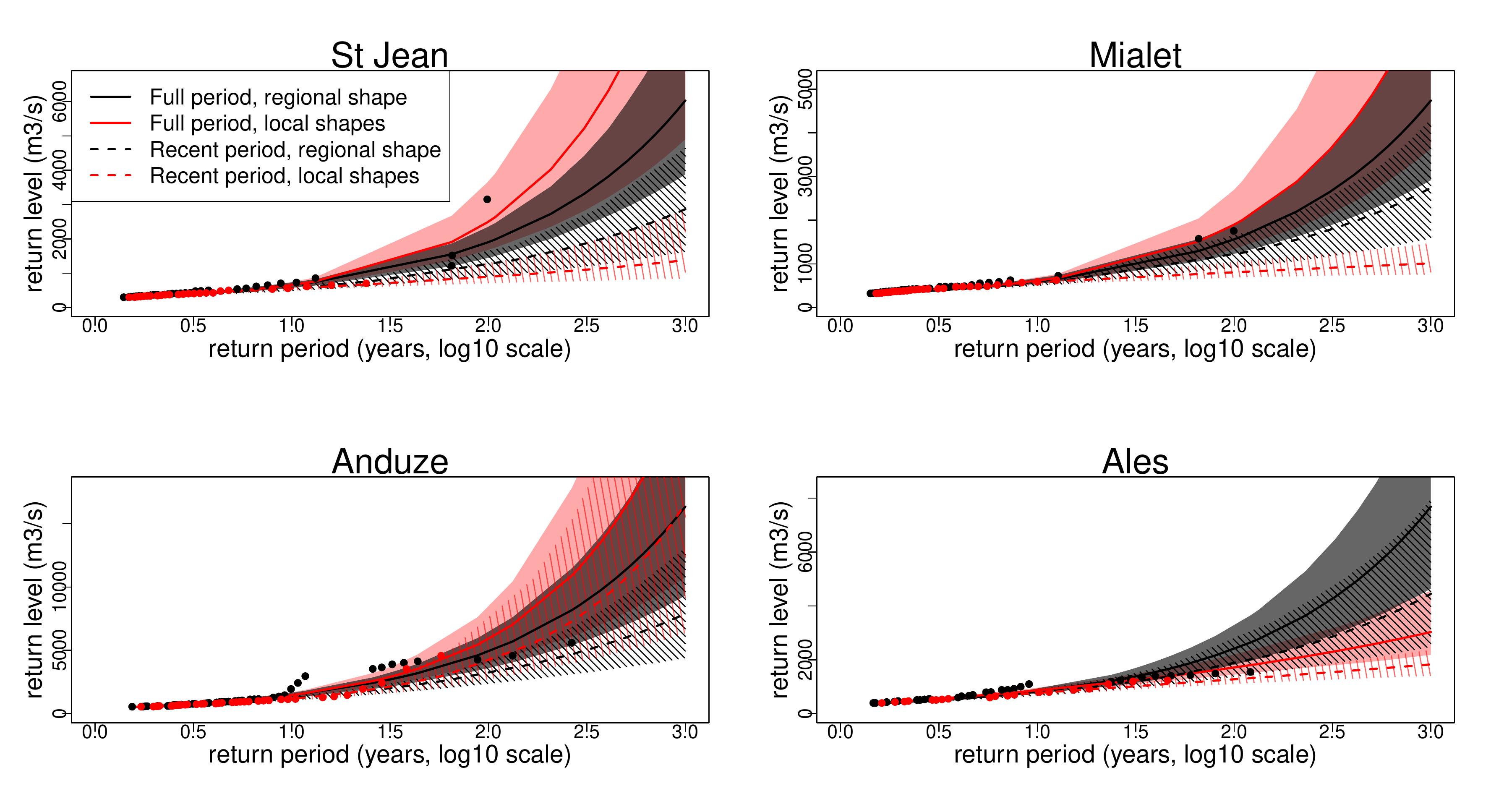}
  \caption{Return level plots at each location using four inferential
    frameworks with $90\%$ posterior quantiles. Dotted lines and hatched areas: data from he recent
    period only; Solid lines and shaded area: Full data set; Black
    lines and Grey area: Regional analysis, global shape parameter;
    Red lines and shaded red area: local shape parameters. Black
    (\emph{resp.} Red) points: observed data plotted at the
    corresponding empirical return period using the whole
    (\emph{resp.} recent) data set.}\label{fig:returnLevels}
\end{figure}

Figure~\ref{fig:returnLevels} shows posterior mean estimates of the return
levels at each location, together with credible intervals based on
posterior $0.05-0.95$ quantiles, in  the four inferential
frameworks. The return levels 
appear to be very sensitive to model choice: overall,
taking into account  the whole period increases the estimated return
levels. In terms of mean estimate, 
the effect of imposing a global shape parameter  varies from one
station to another, as expected. For those return levels, the posterior credibility intervals
seem to depend more on the mean return levels than on the choice of a 
regional or local framework. This  seems at odds with the previous
findings of reduced
intervals for marginal parameters. However, one must note that the
width of return level credibility intervals depends not only on that
of the parameters, but also  on the value of
the mean estimates. Larger parameter estimates involve larger
uncertainty in terms of return levels. 

\begin{figure}[hbtp]
  \centering
  \includegraphics[scale=0.38]{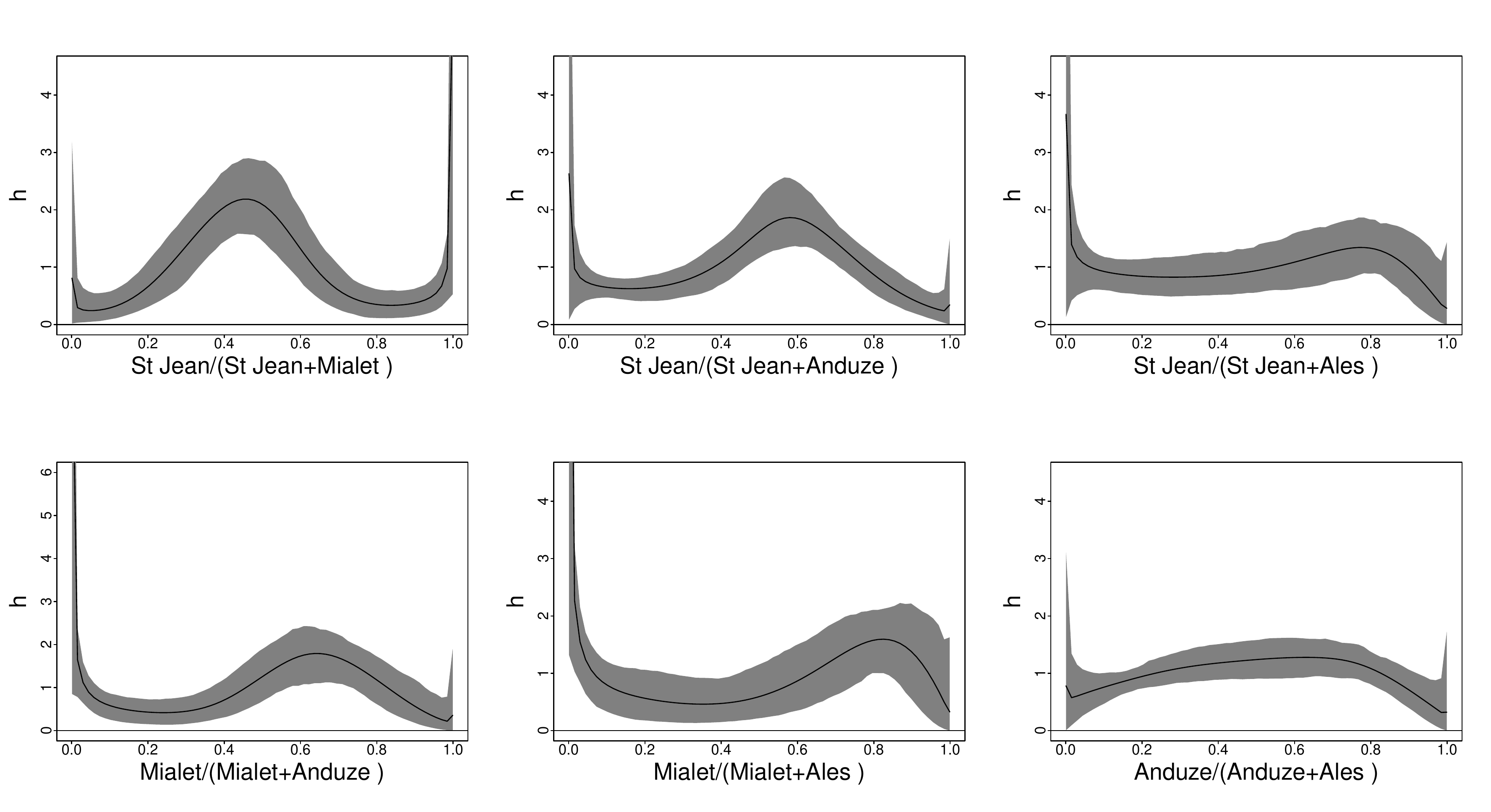}
  \caption{Posterior predictive bi-variate angular densities (black
    lines) with posterior $0.05-0.95$ quantiles (Grey areas). 
}
  \label{fig:angPostPred}
\end{figure}

In addition to uni-variate quantities of interest such as marginal
parameters or return level curves, having estimated the dependence
structure gives access to multivariate quantities. 
Figure~\ref{fig:angPostPred} shows the posterior mean estimates of
the angular density. Since the four-variate version of the angular
distribution cannot be easily represented, the bivariate marginal versions
of the angular distribution are displayed instead. Here, the unit simplex
(which was the diagonal blue segment in Figure~\ref{fig:exampleH}) is
represented by the horizontal axis, so that $H$ is a distribution
function on $[0,1]$. As could be expected in view of
Figure~\ref{fig:bivarPlots}, extremes are rather strongly
dependent. Moreover, the  posterior distribution  is overall well concentrated
around the mean estimate. 

\begin{figure}[hbtp]
  \centering
  \includegraphics[scale=0.38]{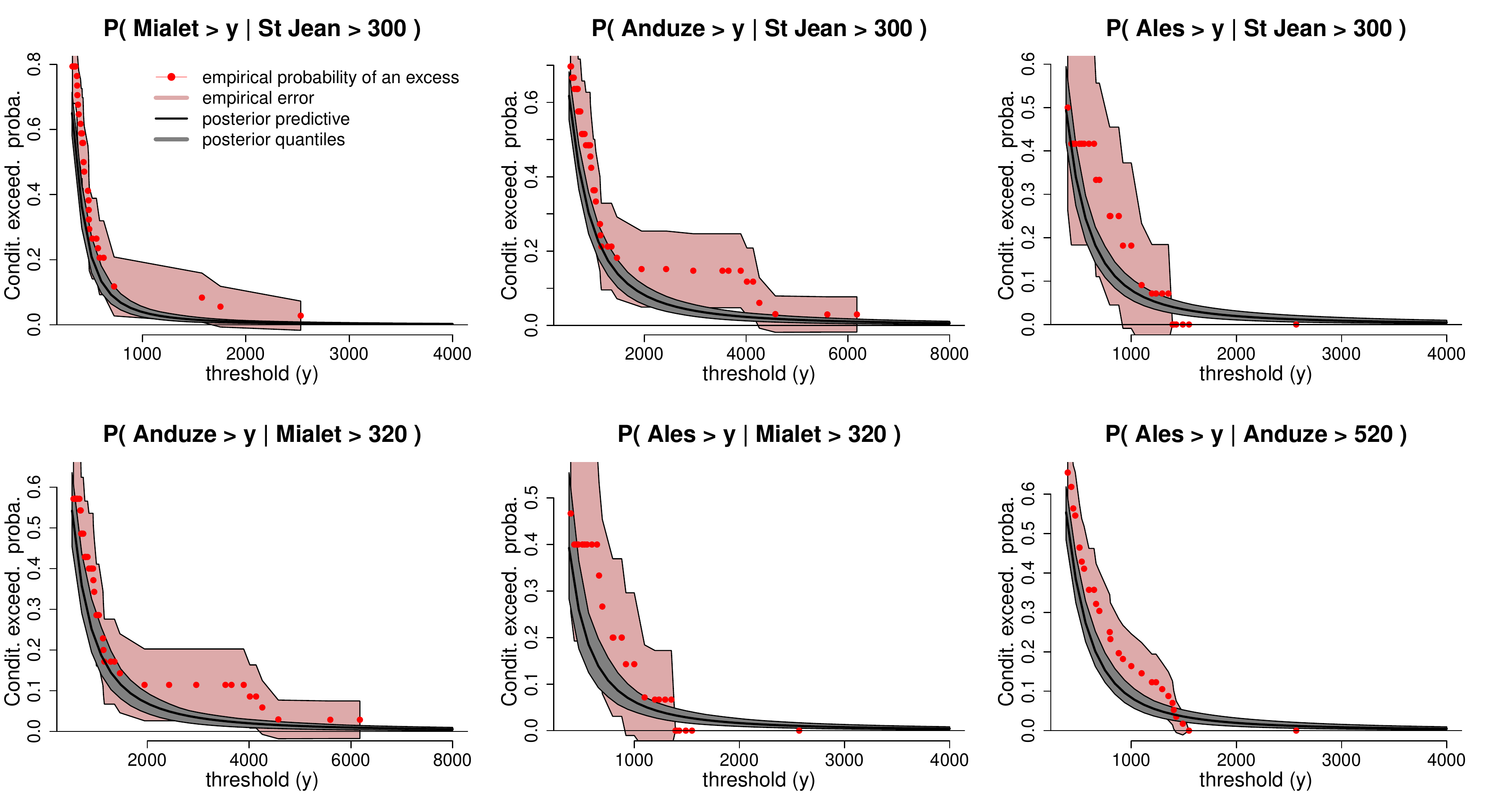}
  \caption{Conditional tail distributions. Black line and Grey area:
    posterior mean estimate and posterior $90\%$ credible intervals
    (posterior quantiles); red
    points: empirical tail function computed at the recorded points
    above threshold; pale red area: $90\%$ Gaussian confidence intervals around
    the empirical estimates.}\label{fig:conditPexc}
\end{figure}

The predictive angular distribution allows to estimate conditional
probabilities of exceedance of high thresholds. As an example,  
figure~\ref{fig:conditPexc} displays, for the six pairs $1\le j<i\le
4$,
the posterior estimates of the   conditional  tail distribution 
functions $P(Y_i^\vee >y |Y_j^\vee > \thres_j)$
at  location $i$,
conditioned upon an excess of the threshold $\thres_j$ at another
location $j$.  The predictive tail
functions  in the
\DM\ model concur with the empirical estimates for moderate values of
$y$. For larger values, the empirical error grows and no empirical
estimate exists outside the observed domain. However, the \DM\ estimates
are still defined and the size of the error region remains
comparatively small. 

\begin{figure}[hbtp]
  \centering
  \includegraphics[scale=0.3]{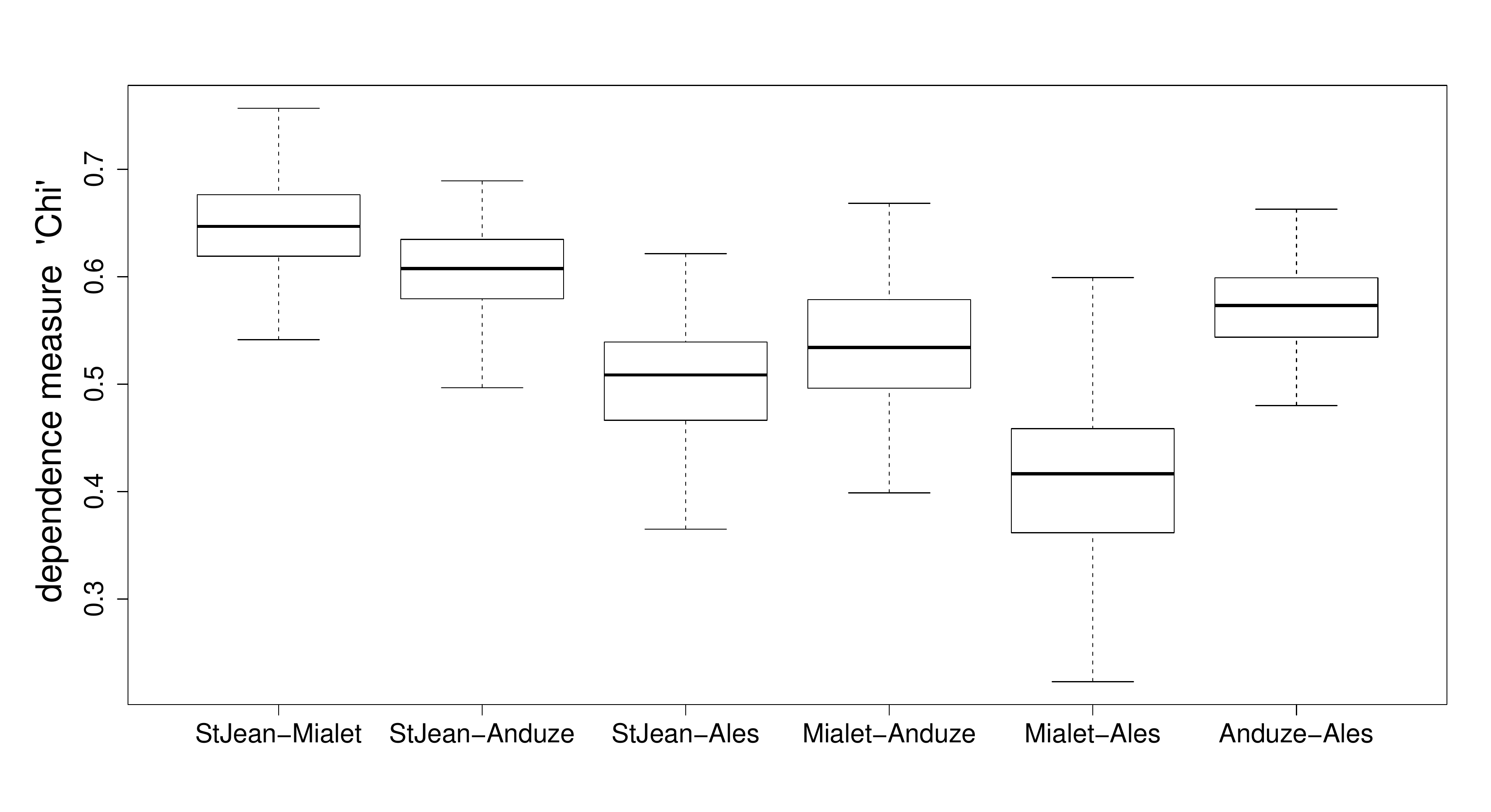}
  \caption{Dependence measure $\boldsymbol{\chi}_{i,j}$ for the six pairs of locations: Posterior box-plot.}
  \label{fig:boxplot_extremalDepCoeff}
\end{figure}

Finally, one commonly used measure of dependence at asymptotically high
levels  between pairs
of locations is  defined by \citep{Coles1999}:

$$
\boldsymbol{\chi}_{i,j} = \lim_{x\to\infty} \frac{P(X_i > x, X_j>x  )}{P(X_j>x)} = 
\lim_{x\to \infty}P(X_i >x \left\vert X_j>x\right.)\,,
$$
where $X_i$, $X_j$ are the Fréchet-transformed variables at locations
$i$ and $j$. Since $X_i$ and $X_j$ are identically distributed,
$\boldsymbol{\chi}_{i,j} = \boldsymbol{\chi}_{j,i}$. 
From its definition, $\boldsymbol{\chi}_{i,j}$ is comprised between $0$ and $1$;
small values indicate weak dependence at high levels whereas values
close to $1$ are characteristic of strong dependence. In the extreme
case $\boldsymbol{\chi}=0$, the variables are asymptotically independent. 
In the case of Dirichlet mixtures, $\boldsymbol{\chi}_{i,j}$ has an explicit
expression formed of incomplete Beta functions
\citep[][eq. (9)]{Boldi_Davison07}. 
Figure~\ref{fig:boxplot_extremalDepCoeff} shows posterior box-plots of
$\boldsymbol{\chi}$ for
the six pairs. The strength of the dependence and the amount of
uncertainty varies from one pair to another, but mean estimates are
overall large (greater than $0.4$), indicating strong asymptotic dependence. 

\begin{figure}[hbtp]
  \centering
  \includegraphics[scale=0.38]{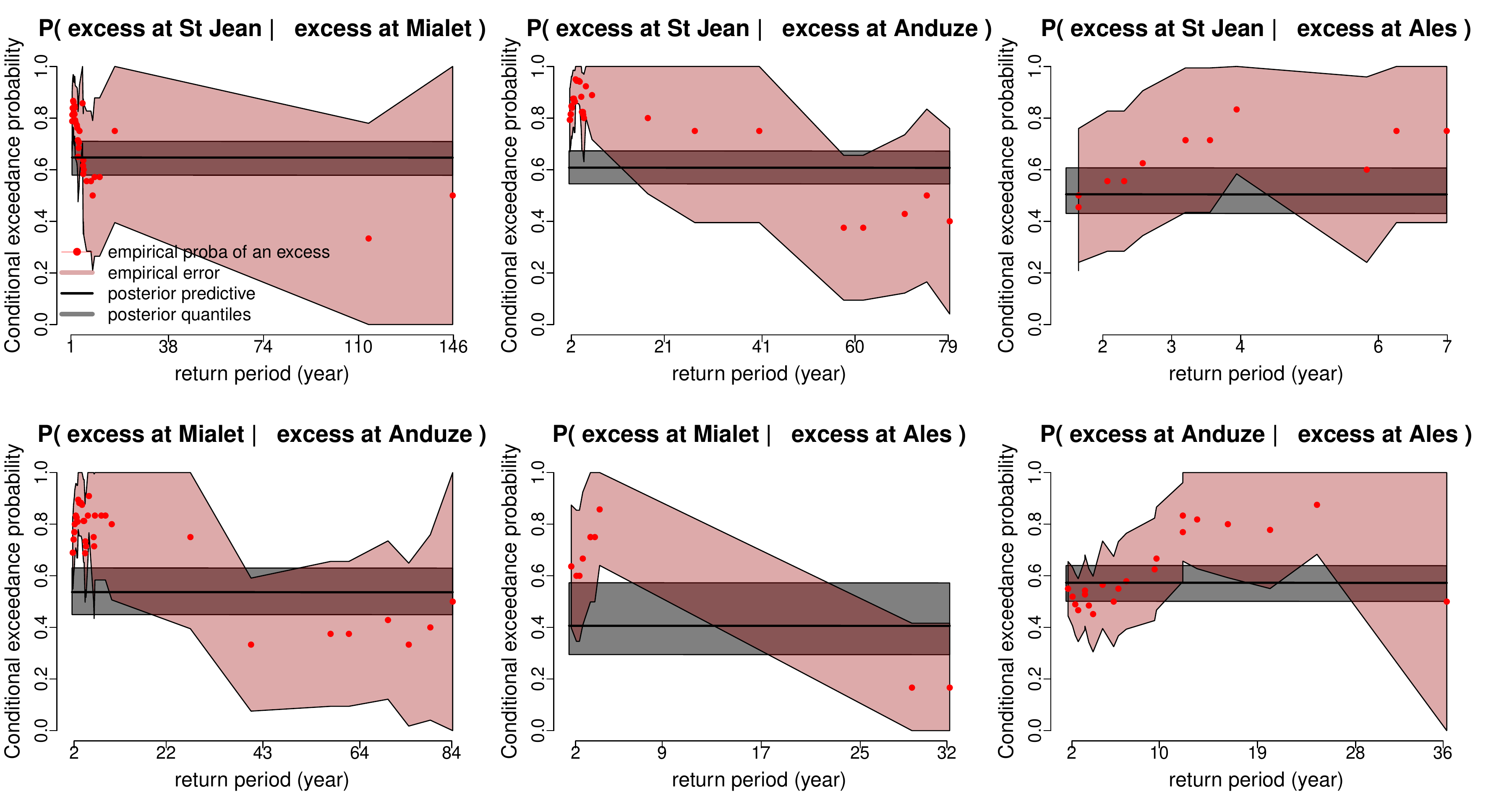}
  \caption{Observed conditional probability of
    exceedance of equally scaled thresholds. Red points: empirical
    estimates of conditional excesses; pale red regions, empirical
    standard error; horizontal black line and gray area, posterior
    mean and $0.05-0.95$ quantiles of the theoretical value 
    in the \DM\ model.}
  \label{fig:extremal_dependence_emp}
\end{figure}

In order to verify the consistency of those results with observed
data, empirical quantities $P( X_i > x |  X_j >x)$ have been
computed and are displayed in Figure~\ref{fig:extremal_dependence_emp}. 
More precisely, it is easy to see that 
$$
P( X_i > x |  X_j >x) = 
P(Y_i^\vee > (F_i^{\margpar})^{-1}\circ F_j^{\margpar}(y) | Y_j^\vee > y)\,,
$$
where $F_i^{\margpar}, F_{j}^{\margpar}$ are the marginal \cdf\ for
location $i$ and $j$, and the $Y_j^\vee,\,Y_i^\vee$'s are the observed
data (cluster maxima). In
Figure~\ref{fig:extremal_dependence_emp}, the 
conditioning thresholds $y$ are the observed values of the
conditioning variable $Y_j^\vee$ above the initial threshold $\thres_j$,
of which the estimated return period (abscissa of the red points) is 
taken  as its   mean estimate using  the
marginal parameter components of the posterior sample.  For each such $y$, $(F_i^{\margpar})^{-1}\circ
F_j^{\margpar}(y)$ is estimated by its posterior mean value, again 
computed
from the marginal posterior sample. Then, the conditional probability  of an
excess by $Y_i^\vee$ (Y-axis value
of the red points) is computed empirically. In theory, as the 
return period increases, the red
points should come closer to the horizontal black line, which is the
mean estimate of $\boldsymbol{\chi}$ computed in the Dirichlet mixture 
dependence model, as in 
Figure~\ref{fig:boxplot_extremalDepCoeff}. Note that in the Dirichlet
model, the limiting value $\boldsymbol{\chi}$ is already reached at finite
levels  because the
conditional probability of an excess on the Fréchet scale, 
$P(X_i>x |X_j >x)$, is constant in $x$. On the contrary, in an asymptoticallly
independent model, the conditional exceeedance probability whould be decreasing
towards zero. 
Results in 
Figure~\ref{fig:extremal_dependence_emp} are comforting: 
the 
mean values of $\boldsymbol{\chi}$ obtained from the Dirichet model
are
within the error regions of the empirical
estimates. The latter  are very large, compared to the posterior
quantiles from the Dirichlet mixture, which illustrates the usefulness
of an extreme value model for computing conditional probabilities of
an excess. 

This result has implications for computing the return periods of
joint excesses of high thresholds. Consider, for example, the  10 years
marginal 
return levels at two  stations, $(q_1, q_2)$. If the excesses above
these threhsolds were assumed to be independent,  taking into account short term temporal
dependence (the mean cluster  size is $\tau = 1.248$),
the return period for the joint excess $(Y_1^\vee>q_1, Y_2^\vee>q_2)$ would be 
$10^2 * (365/\tau) = 29247.8$ years.  
On the contrary, accounting for spatial dependence, 
for example between the two first stations (St Jean and Mialet),
yields an estimated 
return period for a joint excess of 
$10 / \hat{\boldsymbol{\chi}}_{1,2}  = 10/0.645 =   15.5 $ years. 

\section{Discussion}\label{sec:discussion}

This section lists the limitations of the model used in this paper and discusses directions for improvement.

\subsection{Impact of systematic rating curve errors}

The use of historical data allows extending the period of record and hence the availability of extreme flood events.
However, historical data are also usually much more uncertain than recent systematic data, for two reasons:
(i) the precision of historical water stages is limited; (ii) the transformation of these stage values into
discharge values is generally based on a rating curve derived using a hydraulic model, which may induce 
large systematic errors.

The model used in the present paper ignores systematic errors (ii). This is because we focused on multivariate aspects
through the use of the \DM\ model to describe intersite dependence. However, systematic errors may have 
a non-negligible impact on marginal quantile estimates, as discussed by \cite{neppel2010flood}. 
Moreover, in a multivariate context, the impact of systematic errors on the estimation of the 
dependence structure is unclear at this stage and requires further evaluation. Future work will therefore
aim at incorporating an explicit treatment of systematic errors, using models such as those discussed by
\cite{Reis2005} or \cite{neppel2010flood}.

\subsection{Comparing several models for intersite dependence}

The \DM\ model used in this paper to describe intersite dependence is a valid dependence model according to multivariate
extreme value theory (MEVT). Many alternative approaches, not necessarily MEVT-compatible, have been proposed in the 
hydrological literature on regional estimation methods. Such approaches include simply ignoring dependence \citep[e.g.][]{Dalrymple1960}, the concept
of 'equivalent number of sites' \citep{Reed1999} or the use of copulas \citep[e.g.][]{Renard2011}. This raises the question 
of the influence of the approach used to describe dependence on the following estimates:
\begin{itemize}
\item Marginal estimates, typically quantile estimates at each site. While the impact of ignoring dependence altogether has been studied by
  several authors \citep{Stedinger1983,Hosking1988,Madsen1997,Renard2007}, the impact of alternative dependence models is less clear. In particular, 
  since marginal estimates do not directly use the dependence model, it remains to be established whether or not different dependence models
  (e.g. asymptotically dependent vs. asymptotically independent) yield significantly different results.
\item Joint or conditional estimates, as illustrated in
  Figures~\ref{fig:bivarPlots}, \ref{fig:conditPexc} and \ref{fig:extremal_dependence_emp} for instance. The dependence model obviously plays a much more 
  important role in this case.
\end{itemize}
Such comparison has not been attempted in this paper because the use of censored historical data makes the application of standard methods like
copulas much more challenging. 

\subsection{The treatment of intersite dependence in a highly dimensional context}

As illustrated in the case study, the \DM\ model is applicable in
moderate dimension d=4. However, such semi-parametric approach
is not geared toward highly-dimensional contexts (e.g. spatial rainfall using dozens or hundreds of rain gauges, or gridded data sets). Practical approaches for highly-dimensional
multivariate extremes have been mostly proposed in the context of
block maxima, using the theory of max-stable processes
\citep{de1984spectral,smith1990max,schlather2002models,westra2011detection}.
Estimation procedures 
\emph{e.g.} using composite likelihood methods exist for such processes
\citep{padoan2010likelihood}, along with descriptive tools
\emph{e.g.} to  define and 
estimate  extremal dependence coefficients such as the madogram
\citep{cooley2006variograms}. 
However, the development of models adapted to peaks-over-threshold is
still an area of active research in a highly-dimensional spatial
context and 
full modeling (which would \emph{e.g} allow simulation of joint excesses) remain
elusive. Recent theoretical advances \citep{ferreira2012generalized,dombry2013functional} 
give cause to hope for, and expect, future  development of spatial
peaks-over-threshold models. 


\section{Conclusion}\label{sec:conclusion}

This paper illustrates the use of a multivariate peaks-over-threshold
model to combine regional estimation and historical floods. This model
is based on a semi-parametric Dirichlet Mixture to describe intersite
dependence, while Generalized Pareto distributions are used for
margins. A  data augmentation scheme is used to enable the inclusion of censored historical flood data. The model is applied to four catchments in Southern France where historical flood data are available.

The first objective of this case study was to assess the relative impact of regional and
historical information on marginal quantile estimates at each site. The main results can be summarized as follows:
\begin{itemize}
\item Over the four considered versions of the model, the version ignoring historical floods and performing local estimation yields estimates that may strongly differ from the other versions. The three other versions (which either use historical floods or perform regional estimation or both) yield more consistent estimates. This illustrates the benefit of extending the at-site sample using either historical or regional information, or both.
\item Compared with the most complete version of the model (which enables both historical floods and regional estimation), the version only implementing regional estimation (but ignoring historical floods) yields smaller estimates of the shape parameter, and hence smaller quantiles. This result is likely specific to this particular data set, for which many large floods have been recorded during the historical period.
\item Compared with the most complete version of the model, the version using historical floods but implementing local estimation yields higher quantiles for three catchments but lower quantiles on the fourth. 
\item The uncertainty in parameter estimates generally decreases when more information (regional, historical or both) is included in the inference. However, this does not necessarily result in smaller uncertainty in quantile estimates. This is because this uncertainty does not only depends on the uncertainty in parameter estimates, but also on the value taken by the parameters. In particular, a precise but large shape parameter may result in more uncertain quantiles than a more imprecise but lower shape parameter.
\end{itemize}

The second objective was to investigate the nature of asymptotic
dependence in this flood data set, by taking advantage of the existence
of extremely high joint exceedances in the historical data. Results
in terms of predictive angular density 
suggest the existence of  such  dependence between 
every pairs of catchments of asymmetrical nature: some pairs are more
dependent than others at asymptotic levels. In addition,  the Dirichlet Mixture model allows to 
compute bi-variate conditional probabilities of large threshold exceedances,
which are poorly estimated with empirical methods. The limiting values
of the 
conditional probabilities, theoretically obtained with increasing thresholds, are
substantially non zero (they range between $0.4$ and $0.65$), which
confirms the strength and the asymmetry of pairwise  
asymptotic dependence  for this data set and induces multivariate  return
periods much shorter than they would be in the asymptotically
independent case. 
\section{Acknowledgement}
The first author would like to thank Anne-Laure Fougères and
  Philippe Naveau for their useful advice. 
Part of this work has been supported by the EU-FP7 ACQWA Project 
(www.acqwa.ch), by the PEPER-GIS project, by the ANR (MOPERA, McSim,
StaRMIP)  and by the MIRACCLE-GICC project. 




\bibliographystyle{apalike} 
\bibliography{HistFloodsCensoredPOT.bbl}

\begin{thebibliography}{}

\bibitem[Beirlant et~al., 2004]{Beirlant04}
Beirlant, J., Goegebeur, Y., Segers, J., and Teugels, J. (2004).
\newblock {\em Statistics of extremes: Theory and applications}.
\newblock John Wiley \& Sons: New York.

\bibitem[Boldi and Davison, 2007]{Boldi_Davison07}
Boldi, M.-O. and Davison, A.~C. (2007).
\newblock A mixture model for multivariate extremes.
\newblock {\em Journal of the Royal Statistical Society: Series B (Statistical
  Methodology)}, 69(2):217--229.

\bibitem[Coles, 2001]{coles2001introduction}
Coles, S. (2001).
\newblock {\em An introduction to statistical modeling of extreme values}.
\newblock Springer Verlag.

\bibitem[Coles et~al., 1999]{Coles1999}
Coles, S., Heffernan, J., and Tawn, J.~A. (1999).
\newblock Dependence measures for extreme value analyses.
\newblock {\em Extremes}, 2:339--365.

\bibitem[Coles and Tawn, 1991]{coles1991modeling}
Coles, S. and Tawn, J. (1991).
\newblock Modeling extreme multivariate events.
\newblock {\em JR Statist. Soc. B}, 53:377--392.

\bibitem[Cooley et~al., 2006]{cooley2006variograms}
Cooley, D., Naveau, P., and Poncet, P. (2006).
\newblock Variograms for spatial max-stable random fields.
\newblock In {\em Dependence in probability and statistics}, pages 373--390.
  Springer.

\bibitem[Dalrymple, 1960]{Dalrymple1960}
Dalrymple, T. (1960).
\newblock Flood frequency analyses.
\newblock {\em Water-supply paper 1543-A}.

\bibitem[Davison and Smith, 1990]{davison1990models}
Davison, A. and Smith, R. (1990).
\newblock Models for exceedances over high thresholds.
\newblock {\em Journal of the Royal Statistical Society. Series B
  (Methodological)}, pages 393--442.

\bibitem[De~Haan, 1984]{de1984spectral}
De~Haan, L. (1984).
\newblock A spectral representation for max-stable processes.
\newblock {\em The annals of probability}, pages 1194--1204.

\bibitem[De~Haan and De~Ronde, 1998]{DeHaan1998}
De~Haan, L. and De~Ronde, J. (1998).
\newblock Sea and wind: Multivariate extremes at work.
\newblock {\em Extremes}, 1:7--45.

\bibitem[Dombry and Ribatet, 2013]{dombry2013functional}
Dombry, C. and Ribatet, M. (2013).
\newblock Functional regular variations, pareto processes and peaks over
  threshold.

\bibitem[Ferreira and de~Haan, 2012]{ferreira2012generalized}
Ferreira, A. and de~Haan, L. (2012).
\newblock The generalized pareto process; with a view towards application and
  simulation.
\newblock {\em arXiv preprint arXiv:1203.2551v2}.

\bibitem[Ferro and Segers, 2003]{ferro2003inference}
Ferro, C. and Segers, J. (2003).
\newblock Inference for clusters of extreme values.
\newblock {\em Journal of the Royal Statistical Society: Series B (Statistical
  Methodology)}, 65(2):545--556.

\bibitem[Gaume et~al., 2010]{Gaume2010}
Gaume, E., Gaal, L., Viglione, A., Szolgay, J., Kohnova, S., and Bloschl, G.
  (2010).
\newblock Bayesian mcmc approach to regional flood frequency analyses involving
  extraordinary flood events at ungauged sites.
\newblock {\em Journal of Hydrology}, 394:101--117.

\bibitem[Gelman and Rubin, 1992]{gelman1992inference}
Gelman, A. and Rubin, D. (1992).
\newblock Inference from iterative simulation using multiple sequences.
\newblock {\em Statistical science}, pages 457--472.

\bibitem[Gumbel, 1960]{gumbel1960distributions}
Gumbel, E. (1960).
\newblock Distributions des valeurs extr{\^e}mes en plusieurs dimensions.
\newblock {\em Publ. Inst. Statist. Univ. Paris}, 9:171--173.

\bibitem[Heidelberger and Welch, 1983]{heidelberger1983simulation}
Heidelberger, P. and Welch, P. (1983).
\newblock Simulation run length control in the presence of an initial
  transient.
\newblock {\em Operations Research}, pages 1109--1144.

\bibitem[Hosking, 1985]{Hosking1985}
Hosking, J. (1985).
\newblock Maximum-likelihood estimation of the parameters of the generalized
  extreme-value distribution.
\newblock {\em Applied Statistics}, 34:301--310.

\bibitem[Hosking and Wallis, 1987]{Hosking1987}
Hosking, J. and Wallis, J.~R. (1987).
\newblock Parameter and quantile estimation for the generalized pareto
  distribution.
\newblock {\em Technometrics}, 29(3):339--349.

\bibitem[Hosking and Wallis, 1988]{Hosking1988}
Hosking, J. and Wallis, J.~R. (1988).
\newblock The effect of intersite dependence on regional flood frequency
  analysis.
\newblock {\em Water Resources Research}, 24:588--600.

\bibitem[Hosking and Wallis, 1997]{Hosking1997}
Hosking, J. and Wallis, J.~R. (1997).
\newblock {\em Regional Frequency Analysis: an approach based on L-Moments}.
\newblock Cambridge University Press, Cambridge, UK.

\bibitem[Jin and Stedinger, 1989]{Jin1989}
Jin, M. and Stedinger, J.~R. (1989).
\newblock Flood frequency analysis with regional and historical information.
\newblock {\em Water Resources Research}, 25(5):925--936.

\bibitem[Joe et~al., 1992]{joe1992bivariate}
Joe, H., Smith, R.~L., and Weissman, I. (1992).
\newblock Bivariate threshold methods for extremes.
\newblock {\em Journal of the Royal Statistical Society. Series B
  (Methodological)}, pages 171--183.

\bibitem[Katz et~al., 2002]{katz2002statistics}
Katz, R.~W., Parlange, M.~B., and Naveau, P. (2002).
\newblock Statistics of extremes in hydrology.
\newblock {\em Advances in water resources}, 25(8):1287--1304.

\bibitem[Lang et~al., 1999]{Lang1999}
Lang, M., Ouarda, T., and Bobee, B. (1999).
\newblock Towards operational guidelines for over-threshold modeling.
\newblock {\em Journal of Hydrology}, 225:103--117.

\bibitem[Leadbetter, 1983]{leadbetter1983extremes}
Leadbetter, M. (1983).
\newblock Extremes and local dependence in stationary sequences.
\newblock {\em Probability Theory and Related Fields}, 65(2):291--306.

\bibitem[Ledford and Tawn, 1996]{ledford1996statistics}
Ledford, A. and Tawn, J. (1996).
\newblock Statistics for near independence in multivariate extreme values.
\newblock {\em Biometrika}, 83(1):169--187.

\bibitem[Madsen et~al., 1997a]{Madsen1997c}
Madsen, H., Pearson, C.~P., and Rosbjerg, D. (1997a).
\newblock Comparison of annual maximum series and partial duration series
  methods for modeling extreme hydrologic events .2. regional modeling.
\newblock {\em Water Resources Research}, 33(4):759--769.

\bibitem[Madsen et~al., 1997b]{Madsen1997b}
Madsen, H., Rasmussen, P.~F., and Rosbjerg, D. (1997b).
\newblock Comparison of annual maximum series and partial duration series
  methods for modeling extreme hydrologic events .1. at-site modeling.
\newblock {\em Water Resources Research}, 33(4):747--757.

\bibitem[Madsen and Rosbjerg, 1997]{Madsen1997}
Madsen, H. and Rosbjerg, D. (1997).
\newblock The partial duration series method in regional index-flood modeling.
\newblock {\em Water Resources Research}, 33(4):737--746.

\bibitem[Nadarajah, 2001]{nadarajah2001multivariate}
Nadarajah, S. (2001).
\newblock Multivariate declustering techniques.
\newblock {\em Environmetrics}, 12(4):357--365.

\bibitem[Naulet et~al., 2005]{naulet2005flood}
Naulet, R., Lang, M., Ouarda, T.~B., Coeur, D., Bob{\'e}e, B., Recking, A., and
  Moussay, D. (2005).
\newblock Flood frequency analysis on the ard{\`e}che river using french
  documentary sources from the last two centuries.
\newblock {\em Journal of Hydrology}, 313(1):58--78.

\bibitem[Neppel et~al., 2010]{neppel2010flood}
Neppel, L., Renard, B., Lang, M., Ayral, P., Coeur, D., Gaume, E., Jacob, N.,
  Payrastre, O., Pobanz, K., and Vinet, F. (2010).
\newblock Flood frequency analysis using historical data: accounting for random
  and systematic errors.
\newblock {\em Hydrological Sciences Journal--Journal des Sciences
  Hydrologiques}, 55(2):192--208.

\bibitem[O'Connel et~al., 2002]{OConnel2002}
O'Connel, D., Ostenaa, D., Levish, D., and Klinger, R. (2002).
\newblock Bayesian flood frequency analysis with paleohydrologic bound data.
\newblock {\em Water Resources Research}, 38(5).

\bibitem[Padoan et~al., 2010]{padoan2010likelihood}
Padoan, S.~A., Ribatet, M., and Sisson, S.~A. (2010).
\newblock Likelihood-based inference for max-stable processes.
\newblock {\em Journal of the American Statistical Association}, 105(489).

\bibitem[Parent and Bernier, 2003]{Parent2003}
Parent, E. and Bernier, J. (2003).
\newblock Bayesian pot modeling for historical data.
\newblock {\em Journal of hydrology}, 274:95--108.

\bibitem[Payrastre et~al., 2011]{Payrastre2011}
Payrastre, O., Gaume, E., and Andrieu, H. (2011).
\newblock Usefulness of historical information for flood frequency analyses:
  Developments based on a case study.
\newblock {\em Water Resources Research}, 47.

\bibitem[Reed et~al., 1999]{Reed1999}
Reed, D.~W., Faulkner, D.~S., and Stewart, E.~J. (1999).
\newblock The forgex method of rainfall growth estimation - ii: Description.
\newblock {\em Hydrology and Earth System Sciences}, 3(2):197--203.

\bibitem[Reis and Stedinger, 2005]{Reis2005}
Reis, D. and Stedinger, J.~R. (2005).
\newblock Bayesian mcmc flood frequency analysis with historical information.
\newblock {\em Journal of Hydrology}, 313(1-2):97--116.

\bibitem[Renard, 2011]{Renard2011}
Renard, B. (2011).
\newblock A bayesian hierarchical approach to regional frequency analysis.
\newblock {\em Water Resources Research}, 47.

\bibitem[Renard and Lang, 2007]{Renard2007}
Renard, B. and Lang, M. (2007).
\newblock Use of a gaussian copula for multivariate extreme value analysis:
  some case studies in hydrology.
\newblock {\em Advances in Water Resources}, 30(4):897--912.

\bibitem[Resnick, 1987]{resnick1987extreme}
Resnick, S. (1987).
\newblock {\em Extreme values, regular variation, and point processes, volume 4
  of Applied Probability. A Series of the Applied Probability Trust}.
\newblock Springer-Verlag, New York.

\bibitem[Resnick, 2007]{Resnick07}
Resnick, S. (2007).
\newblock {\em Heavy-Tail Phenomena: Probabilistic and Statistical Modeling}.
\newblock Springer Series in Operations Research and Financial Engineering.

\bibitem[Ribereau et~al., 2011]{ribereau2011note}
Ribereau, P., Naveau, P., and Guillou, A. (2011).
\newblock A note of caution when interpreting parameters of the distribution of
  excesses.
\newblock {\em Advances in Water Resources}, 34(10):1215--1221.

\bibitem[Sabourin, 2014]{sabourin14censoring}
Sabourin, A. (2014).
\newblock Semi-parametric modeling of excesses above high multivariate
  thresholds with censored data.
\newblock {\em submitted}.

\bibitem[Sabourin and Naveau, 2013]{sabourinNaveau2012}
Sabourin, A. and Naveau, P. (2013).
\newblock Bayesian dirichlet mixture model for multivariate extremes: A
  re-parametrization.
\newblock {\em Computational Statistics \& Data Analysis}, DOI
  10.1016/j.csda.2013.04.021.

\bibitem[Schlather, 2002]{schlather2002models}
Schlather, M. (2002).
\newblock Models for stationary max-stable random fields.
\newblock {\em Extremes}, 5(1):33--44.

\bibitem[Smith, 1994]{smith1994multivariate}
Smith, R. (1994).
\newblock Multivariate threshold methods.
\newblock {\em Extreme Value Theory and Applications}, 1:225--248.

\bibitem[Smith et~al., 1997]{smith1997markov}
Smith, R., Tawn, J., and Coles, S. (1997).
\newblock Markov chain models for threshold exceedances.
\newblock {\em Biometrika}, 84(2):249--268.

\bibitem[Smith, 1990]{smith1990max}
Smith, R.~L. (1990).
\newblock Max-stable processes and spatial extremes.
\newblock {\em Unpublished manuscript, Univer}.

\bibitem[Stedinger, 1983]{Stedinger1983}
Stedinger, J.~R. (1983).
\newblock Estimating a regional flood frequency distribution.
\newblock {\em Water Resources Research}, 19:503--510.

\bibitem[Stedinger and Cohn, 1986]{Stedinger1986}
Stedinger, J.~R. and Cohn, T.~A. (1986).
\newblock Flood frequency-analysis with historical and paleoflood information.
\newblock {\em Water Resources Research}, 22(5):785--793.

\bibitem[Stephenson, 2003]{stephenson2003simulating}
Stephenson, A. (2003).
\newblock Simulating multivariate extreme value distributions of logistic type.
\newblock {\em Extremes}, 6(1):49--59.

\bibitem[Stephenson, 2009]{stephenson2009high}
Stephenson, A. (2009).
\newblock High-dimensional parametric modelling of multivariate extreme events.
\newblock {\em Australian \& New Zealand Journal of Statistics}, 51(1):77--88.

\bibitem[Tanner and Wong, 1987]{tanner1987calculation}
Tanner, M. and Wong, W. (1987).
\newblock The calculation of posterior distributions by data augmentation.
\newblock {\em Journal of the American Statistical Association},
  82(398):528--540.

\bibitem[Tasker and Stedinger, 1987]{Tasker1987}
Tasker, G.~D. and Stedinger, J.~R. (1987).
\newblock Regional regression of flood characteristics employing historical
  information.
\newblock {\em Journal of Hydrology}, 96:255--264.

\bibitem[Tasker and Stedinger, 1989]{Tasker1989}
Tasker, G.~D. and Stedinger, J.~R. (1989).
\newblock An operational gls model for hydrologic regression.
\newblock {\em Journal of Hydrology}, 111:361:375.

\bibitem[Van~Dyk and Meng, 2001]{van2001art}
Van~Dyk, D. and Meng, X. (2001).
\newblock The art of data augmentation.
\newblock {\em Journal of Computational and Graphical Statistics}, 10(1):1--50.

\bibitem[Westra and Sisson, 2011]{westra2011detection}
Westra, S. and Sisson, S.~A. (2011).
\newblock Detection of non-stationarity in precipitation extremes using a
  max-stable process model.
\newblock {\em Journal of Hydrology}, 406(1):119--128.

\end{thebibliography}

\end{document}